\documentclass[prb,a4paper,floatfix,superscriptaddress,secnumarabic,twocolumn]{revtex4}

\usepackage[english]{babel}
\usepackage{amsmath,amssymb}
\usepackage{stmaryrd,dsfont,txfonts}
\usepackage{subfigure}
\usepackage{graphicx}
\usepackage{dcolumn}
\usepackage[sort&compress]{natbib}
\usepackage{float}
\usepackage{ifpdf}
\ifpdf
  \usepackage[dvipdfm,colorlinks,hyperindex]{hyperref}
\else
  \usepackage[hypertex,colorlinks,hyperindex]{hyperref}
\fi

\renewcommand{\arraystretch}{1.5}
\newcommand{\e}{\mathrm{e}}

\newcommand{\eem}{m^{\ast}}

\newcommand{\mce}{\mathcal{E}}
\newcommand{\rpab}{\mathbf{r}_{\shortparallel}}
\newcommand{\kpab}{\mathbf{k}_{\shortparallel}}

\newcommand{\kpa}{k_{\shortparallel}}

\newcommand{\bra}[1]{\ensuremath{\bigl\langle #1 |}}
\newcommand{\ket}[1]{\ensuremath{| #1 \bigr\rangle}}

\newcommand{\kets}[2]{\ensuremath{| #1 \bigr\rangle \otimes | #2\bigr\rangle}}

\newcommand{\ketd}[2]{\ensuremath{| #1,#2 \bigr\rangle}}
\newcommand{\ep}{\epsilon}

\newcommand{\kp}{\ensuremath{\mathbf{k}\cdot\mathbf{p}} }
\newcommand{\bfs}[1]{\ensuremath{\boldsymbol{#1}} }

\newcommand{\bfl}[1]{\mathbf{#1}}
\newcommand{\ti}[1]{\tilde{#1}}
\newcommand{\imsizeA}{0.475\columnwidth}
\newcommand{\imsizeB}{0.950\columnwidth}
\newcommand{\imsizeC}{0.850\columnwidth}

\newcommand{\imsizeF}{0.920\columnwidth}

\begin{document}

\title{Intersubband-induced spin-orbit interaction in quantum wells}
\author{Rafael S. Calsaverini}
\author{Esmerindo Bernardes}
\email{sousa@if.sc.usp.br}
\author{J. Carlos Egues}
\email{egues@if.sc.usp.br}
\affiliation{Instituto de F{\'{\i}}sica de S{\~{a}}o Carlos,
  Universidade de S{\~{a}}o Paulo, 13560-970 S{\~{a}}o Carlos, SP, Brazil}
\author{Daniel Loss}
\affiliation{Department of Physics, University of Basel, CH-4056
Basel, Switzerland}

\date{\today}

\begin{abstract}
  Recently, we have found an additional spin-orbit (SO) interaction in
  quantum wells with two subbands [Bernardes \textit{et al.}, Phys.
  Rev. Lett. \textbf{99}, 076603 (2007)]. This new SO term is
  non-zero even in symmetric geometries, as it arises from the
  intersubband coupling between confined states of distinct parities,
  and its strength is comparable to that of the ordinary Rashba.
  Starting from the $8 \times 8$ Kane model, here we present a
  detailed derivation of this new SO Hamiltonian and the corresponding
  SO coupling.  In addition, within the self-consistent Hartree
  approximation, we calculate the strength of this new SO coupling for
  realistic symmetric modulation-doped wells with two subbands. We
  consider gated structures with either a constant areal electron
  density or a constant chemical potential. In the parameter range
  studied, both models give similar results. By considering the
  effects of an external applied bias, which breaks the structural
  inversion symmetry of the wells, we also calculate the strength of
  the resulting induced Rashba couplings within each subband.
  Interestingly, we find that for double wells the Rashba couplings
  for the first and second subbands interchange signs abruptly across
  the zero bias, while the intersubband SO coupling exhibits a
  resonant behavior near this symmetric configuration.  For
  completeness we also determine the strength of the Dresselhaus
  couplings and find them essentially constant as function of the
  applied bias.
\end{abstract}

\maketitle

\section{Introduction}

The coupling between spatial and spin degrees of freedom in
semiconductors provides an interesting possibility for coherently
manipulating the electron spin via its orbital (charge) motion. For
instance, the proposal of Datta and Das\cite{apl56sd1990b} for a spin
field-effect transistor highlights the use of the spin-orbit (SO)
interaction of Rashba,\cite{spss2eir1960,jpc17yab1984,jetpl39yab1984}
which is electrically tunable,\cite{prb55ge1997,prl78jn1997} to
control -- via spin rotation -- the flow of electrons between
ferromagnetic source and drain.

In addition to the Rashba SO coupling present in heterostructures with
structural inversion asymmetry in the confining potential, there is
the Dresselhaus SO interaction\cite{pr100gd1955} present in both bulk
and confined structures with inversion asymmetry in the underlying
crystal lattice. These spin orbit interactions have played an
important role in the exciting field of semiconductor spintronics as
they underlie a number of interesting physical phenomena and potential
spintronic
applications.\cite{science294saw2001,awschalom2002,rmp76iz2004} For
instance, the effective \textit{zitterbewegung} of spin-polarized wave
packets injected into SO coupled two-dimensional (2D) electron gases
is a very interesting possibility.\cite{prl64js2005,prb73js2006} The
interplay of the Rashba and Dresselhaus interactions can give rise to
conserved spin-rotation symmetries\cite{prl90js2003,prl97bab2006}
relevant for devising robust SO-based devices operating in the
nonballistic regime.\cite{prl90js2003}

Recently, a new type of SO interaction arising in quantum confined
systems with two subbands has been found \cite{prl99eb2007}. Unlike
the usual Rashba SO, this new SO term is nonzero even in wells with
full structural inversion symmetry (and hence it does not produce spin
splitting). This essentially follows from the distinct parities of the
confined states (even and odd), which can couple via the derivative of
a symmetric potential. This intersubband-induced SO coupling is
quadratic in the crystal momentum, unlike the Rashba and the
(linearized) Dresselhaus terms in wells.\cite{prl98jjk2007} As shown
in Ref.~\onlinecite{prl99eb2007}, this SO coupling can give rise to an
unusual \textit{zitterbewegung} (both in position and in
spin\cite{pssc3esb2006}) and a nonzero spin Hall conductivity.

Here we complement and extend the work of
Ref.~\onlinecite{prl99eb2007}: (i) We present a more thorough
derivation of the intersubband-induced SO interaction, starting from
the $8\times 8$ Kane model\cite{jpcs1eok1957,bastard1988,winkler2003}
within the \kp approach. We also slightly generalize the derivation
for confined systems with more than two subbands and
\emph{structurally asymmetric} potentials in which the usual
Rashba-type SO interaction is present. (ii) We perform a detailed
investigation of the relevant SO couplings via a self-consistent
scheme where we solve both Poisson and Schr{\"{o}}dinger equations
numerically (Numerov method) within the Hartree approximation. We
consider realistic modulation-doped single and double quantum wells
with applied external biases, which can change the spatial symmetry of
the wells, and having either a constant areal electron density or a
constant chemical potential.

Our simulations focus on wells with two subbands. For nonzero applied
biases we calculate not only the intersubband-induced SO coupling
$\eta$ but also the Rashba-type couplings $\alpha_0$, $\alpha_1$ for
the first and second subbands, respectively. For completeness, we also
calculate the linearized Dresselhaus SO couplings for each
subband.\cite{prb72jmj2005} For both the constant density and constant
chemical-potential models considered, we find sizable values of the
intersubband SO coupling $\eta$ as compared to the usual Rashba and
Dresselhaus couplings.  Interestingly, for double wells near the
symmetric (zero-bias) configuration we find that $\eta$ has a resonant
behavior, changing its magnitude by a factor of 10. On the other
hand, the Rashba couplings for the first and second subbands abruptly
change signs around the zero-bias voltage. The Dresselhaus couplings
do not show any noticeable behavior around this point, being
essentially constant as a function of the applied bias.

We note that the SO coupling constants $\eta$, $\alpha_0$, and
$\alpha_1$ contain contributions from the potential well (and barrier)
offsets, the electronic Hartree potential, and the external gate plus
the modulation doping potentials. For the \textit{single wells}
investigated here, the external gate (+ modulation doping) is the
dominant contribution to $\alpha_0$ and $\alpha_1$, while $\eta$ is
mostly determined by the Hartree and structural offset contributions.
On the other hand, for the \textit{double wells} studied the
electronic Hartree potential is the dominant contribution to $\eta$,
$\alpha_0$, and $\alpha_1$. Interestingly, the Hartree potential in
this case is highly influenced by the external gate, particularly
around the symmetric (zero-bias) configuration, as the electrons can
localize in either well for small (positive or negative) changes in
the gate potential. This renders $\eta$, $\alpha_0$, and $\alpha_1$
more amenable to gate modulations in double wells as compared to
single wells. Next we outline our work.

In Sec.~\ref{sec:kp} we review the \kp approach and the Kane model.
In Sec.~\ref{sec:so} we present a detailed derivation of our effective
Hamiltonian for electrons in heterostructures with many confined
states within the Kane model. In this section we also show the
relevant expressions for the new intersubband-induced SO coupling
$\eta$ and those for the Rashba $\alpha$ (and Dresselhaus $\beta$) SO
couplings as well. In Sec.~\ref{sec:systems} we describe the quantum
wells investigated and (briefly) the standard self-consistent
calculation performed. We present our results and discussions in
Sec.~\ref{sec:results}. In this section we focus specifically on
realistic single and double-well systems.  Section~\ref{sec:fim}
summarizes our work. In Appendices \ref{ap:sc-ap} and \ref{apx:cis} we
show details of our self-consistent scheme to solve the relevant
Schr{\"{o}}dinger and Poisson equations.

\section{\kp approach and Kane Model}
\label{sec:kp}

Here we briefly review the \kp approach and use it to obtain the $8
\times 8$ Kane model relevant for our derivation of the new
intersubband-induced SO coupling.\cite{bastard1988,winkler2003}

\subsection{Basics of the \kp method}
The single-particle Hamiltonian for an electron with momentum
\bfs{p} in a periodic potential\cite{bastard1988,prb53dmw1996}
$V(\bfs{r})$ with SO is
\begin{equation}
  \label{eq:Hkp}
  H = \frac{\bfs{p}^2}{2m_0} + V(\bfs{r}) + \frac{\hbar}{4m_0^2 c^2}
  \bfs{\sigma}\times\bfs{\nabla} V(\bfs{r}) \cdot \bfs{p},
\end{equation}
where $m_0$ is the bare electron mass and \bfs{\sigma} is a vector
operator defined in terms of the Pauli matrices. With the help of
Bloch's theorem $\psi_{n\bfs{k}}(\bfs{r}) = \exp(i\bfs{k} \cdot
\bfs{r})\, u_{n\bfs{k}}(\bfs{r})$ [$u_{n\bfs{k}}(\bfs{r})$ has the
periodicity of the underlying Bravais lattice] we can rewrite the
Schr{\"{o}}dinger's equation $H\psi_{n\bfs{k}} =
\varepsilon_{n\bfs{k}} \psi_{n\bfs{k}}$, where $n$ indexes the
distinct solutions for each $\bfs{k}$ vector,  in the form
\begin{equation}
  \label{eq:Seq}
  \left[H(\bfs{k}=\bfs{0}) +  W(\bfs{k}) \right]u_{n\bfs{k}}(\bfs{r}) =
  \left(\varepsilon_{n\bfs{k}} -
    \frac{\hbar^2 k^2}{2m_0}\right)u_{n\bfs{k}}(\bfs{r}),
\end{equation}
with
\begin{align}
  \label{eq:Hk0}
  H(\bfs{k}=\bfs{0}) &= -\frac{\hbar^2}{2m_0}\nabla^{2} + V(\bfs{r}) +
  \frac{\hbar}{4m_0^2 c^2}
  \bfs{\sigma}\times\bfs{\nabla} V(\bfs{r}) \cdot \bfs{p}, \\
  \label{eq:pert}
   W(\bfs{k}) &= \frac{\hbar}{m_0}\bfs{k} \cdot \left( \bfs{p}
    + \frac{\hbar}{4m_0 c^2} \bfs{\sigma} \times \bfs{\nabla}
    V(\bfs{r})\right).
\end{align}

As usual, to solve Eq.~\eqref{eq:Seq}, we expand $u_{n\bfs{k}}$ in
terms of the eigenstates $u_{l\bfs{0}}$ at $\bfs{k} = \bfs{0}$
[\textit{i.e.}, $W(\bfs{k}=\bfs{0})=0$] obtained from
\begin{equation}
  \label{eq:seqk0}
  H(\bfs{k}=\bfs{0})\, u_{l\bfs{0}}(\bfs{r}) =
  \varepsilon_{l\bfs{0}}\, u_{l\bfs{0}}(\bfs{r}),
\end{equation}
where $l=1,2,\ldots N$ (in principle, $N\rightarrow \infty$) indexes
the discrete set of levels at $\bfs{k}=\bfs{0}$ [note that
Eq.~\eqref{eq:seqk0} contains the SO interaction, even though
$W(\bfs{0})=0$]. Substituting
\begin{equation}
  \label{eq:expu0}
  u_{n\bfs{k}}(\bfs{r})=\sum_{l=1}^{N} a_{nl}(\bfs{k})\,
  u_{l\bfs{0}}(\bfs{r}),
\end{equation}
into Eq.~\eqref{eq:Seq} and projecting the resulting expression onto
the $u_{l'\bfs{0}}(\bfs{r})$ eigenstate, we find \cite{bastard1988}
\begin{multline}
  \label{eq:kpbulk}
  \sum_{l=1}^{N} \Big[ \big(\varepsilon_{l\bfs{0}} -
  \varepsilon_{n\bfs{k}} + \frac{\hbar^2 k^2}{2m_0}\big) \delta_{ll'} \\+
  \bra{l'}\frac{\hbar}{m_0} \bfs{k}\cdot\bfs{p} +
  \frac{\hbar^{2}}{4m_0^2 c^2}
  \bfs{k}\cdot\bfs{\sigma}\times\bfs{\nabla}V(\bfs{r}) \ket{l}
  \Big]\, a_{nl}(\bfs{k}) = 0.
\end{multline}
Here we use the notation
$\langle\bfs{r}\ket{l}=u_{l\bfs{0}}(\bfs{r})$ and define
\begin{equation}
  \label{eq:All}
  \bra{l'}A\ket{l} = \int d^{3}r\;
  u_{l'\bfs{0}}^{\ast}\,A\,u_{l\bfs{0}},
\end{equation}
with A denoting a Hermitian operator.

\subsection{$8\times 8$ Kane model - bulk case}
\label{sec:kanebulk}
As usual, in order to solve Eq.~\eqref{eq:kpbulk} we have to truncate
the basis set by considering a finite number $N$ of zone-center basis
functions ${u_{l\bfs{0}}(\bfs{r})}$. In addition, since the
$\bfs{k}=\bfs{0}$ Hamiltonian [Eq.~\eqref{eq:Hk0}] contains a SO term,
it is convenient to choose linear combination of basis functions which
are eigenstates of the total angular momentum
$\mathbf{J}=\mathbf{L}+\mathbf{S}$, and its $z$ component $J_z$; here
$\mathbf{L}$ and $\mathbf{S}$ denote the orbital and spin angular
momenta, respectively. In II-VI and III-V (both zincblend) compounds
the relevant conduction and valence bands arise from the ``bonding''
$p$-type and ``anti-bonding'' $s$-type states, respectively. Following
the notation of Refs.~\onlinecite{bastard1988} and
\onlinecite{ivchenko1997}, we summarize in Table \ref{tab:basis} the
set of eight zone-center wave functions we consider here (the kets $|J
J_z\rangle$ are also shown), which are the eigenstates of the
zone-center Schr{\"{o}}dinger's equation \eqref{eq:seqk0} for the
periodic part of the Bloch function. Note that we use the standard
state vector notation $|S\rangle$, $|X\rangle$, $|Y\rangle$, and
$|Z\rangle$ to denote the symmetry of the corresponding ``atomic
orbitals'' (tight-binding view).

Using the ordered basis states $u_1,\ldots,u_8$ in Table
\ref{tab:basis} we can easily write out the matrix Hamiltonian
[Eq.~\eqref{eq:kpbulk}]\cite{bastard1988,prb47td1993,winkler2003}
\begin{widetext}
  \begin{equation}
    \label{eq-matrix}
    H_{8\times 8} =\left[
      \begin{array}{cccccccc}
        \frac{\hbar^2 k^2}{2m_0} & 0 & -\frac{1}{\sqrt{2}}Pk_{+} &
        \sqrt{\frac{2}{3}}Pk_{z} & \frac{1}{\sqrt{6}}Pk_{-} & 0 &
        -\frac{1}{\sqrt{3}}Pk_{z} & -\frac{1}{\sqrt{3}}Pk_{-} \\
        0 & \frac{\hbar^2 k^2}{2m_0} & 0 & -\frac{1}{\sqrt{6}}Pk_{+} &
        \sqrt{\frac{2}{3}}Pk_{z} & \frac{1}{\sqrt{2}}Pk_{-} &
        -\frac{1}{\sqrt{3}}Pk_{+} & \frac{1}{\sqrt{3}}Pk_{z} \\
        -\frac{1}{\sqrt{2}}Pk_{-} & 0 & \frac{\hbar^2 k^2}{2m_0} -
        E_{g} & 0 & 0 & 0 & 0 & 0 \\
        \sqrt{\frac{2}{3}}Pk_{z} & -\frac{1}{\sqrt{6}}Pk_{-} & 0 &
        \frac{\hbar^2 k^2}{2m_0}-E_{g} & 0 & 0 & 0 & 0 \\
        \frac{1}{\sqrt{6}}Pk_{+} & \sqrt{\frac{2}{3}}Pk_{z} & 0 & 0 &
        \frac{\hbar^2 k^2}{2m_0} - E_{g} & 0 & 0 & 0 \\
        0 & \frac{1}{\sqrt{2}}Pk_{+} & 0 & 0 & 0 & \frac{\hbar^2
        k^2}{2m_0} - E_{g} & 0 & 0 \\
      -\frac{1}{\sqrt{3}}Pk_{z} & -\frac{1}{\sqrt{3}}Pk_{-} & 0 & 0 &
        0 & 0 & \frac{\hbar^2 k^2}{2m_0}-E_{g}-\Delta_g  & 0 \\
        -\frac{1}{\sqrt{3}}Pk_{+} & \frac{1}{\sqrt{3}}Pk_{z} & 0 & 0 &
        0 & 0 & 0 & \frac{\hbar^2 k^2}{2m_0}-E_{g}-\Delta_g.
      \end{array}
    \right]
  \end{equation}
\end{widetext}
where $P$ is the usual Kane matrix element\cite{jpcs1eok1957}
\begin{equation}
  \label{eq:Pkane}
  P = - i\frac{\hbar}{m_{0}} \bra{S}p_{x}\ket{X} =
  \hbar\sqrt{\frac{E_{P}}{2m_{0}}},
\end{equation}
expressed in terms of the parameter $E_{P}$
(Ref.~\onlinecite{jap89iv2001}) and $k_\pm=k_x\pm i k_y$.  We have
also used that
$\bra{S}p_{x}\ket{X}=\bra{S}p_{y}\ket{Y}=\bra{S}p_{z}\ket{Z}$.
Equation \eqref{eq-matrix}) is the $8\times 8$ Kane
Hamiltonian\cite{jpcs1eok1957} describing the \textit{s}-type
conduction and \textit{p}-type valence bands around the $\Gamma$ point
in zincblend compounds. Note that the diagonal elements in Hamiltonian
\eqref{eq-matrix} correspond to the eigenenergies
$\varepsilon_{l\bfs{0}}$ of Eq.~\eqref{eq:seqk0}:
$\varepsilon_{1\bfs{0}}=\varepsilon_{2\bfs{0}}=0$ (``conduction-band
states,'' defined as the zero of energy), $\varepsilon_{3\bfs{0}}=
\varepsilon_{4\bfs{0}}= \varepsilon_{5\bfs{0}}=
\varepsilon_{6\bfs{0}}=-E_g$ (``heavy'' and ``light'' hole bands), and
$\varepsilon_{7\bfs{0}}=\varepsilon_{8\bfs{0}}=-E_g-\Delta_g$
(``split-off'' hole band). Here,
\begin{equation}
\Delta_g =\frac{3\hbar^2}{4m_0^2c^2}\bra{X}\frac{\partial
V}{\partial y} \frac{\partial}{\partial x} -\frac{\partial
V}{\partial x}\frac{\partial}{\partial y}\ket{Y} \label{Delta_g}
\end{equation}
is the ``atomic'' SO parameter defining the split-off gap; see
Fig.~\ref{fig:bands4a}, which schematically shows the conduction and
valence bands of a zincblend structure. The circles indicate the
$\bfs{k}=\bfs{0}$ eigenenergies.

The Kane model treats exactly the conduction-valence band couplings
within the truncated set of eight band-edge wave functions. It is
important to emphasize that we have neglected contributions from the
$\bfs{k}$-dependent SO term in Eq.~\eqref{eq:kpbulk}, when
constructing the Kane Hamiltonian
\eqref{eq-matrix}.\cite{jpcs1eok1957} The SO interaction is
accounted for only within the zone center Schr{\"{o}}dinger's equation
\eqref{eq:seqk0} (parameter $\Delta_g$ above). The diagonalization
of the Kane Hamiltonian gives the dispersions
$\varepsilon_{n,\bfs{k}}$ around the $\Gamma$ point. It is known
that the Kane model presented here is not accurate for valence bands
\cite{bastard1988} (\textit{e.g.}, wrong sign of the heavy hole
masses). However, it provides a simplified and accurate description
for the conduction electrons, which is the focus of our work. Next
we discuss the Kane model in the context of heterostructures.

\subsection{Kane model for quantum wells}

\begin{figure}[h]
  \centering
  \subfigure[\label{fig:bands4a} Energy band structure.]
    {\includegraphics{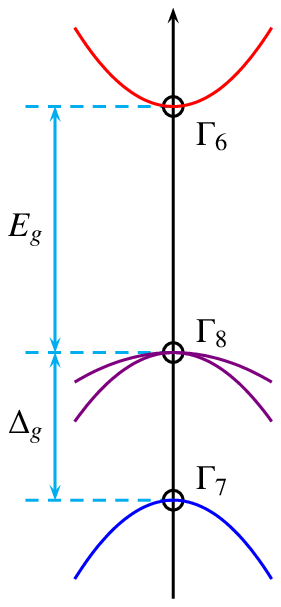}} \quad\quad\quad
  \subfigure[\label{fig:bands4b} Band offsets.]
    {\includegraphics{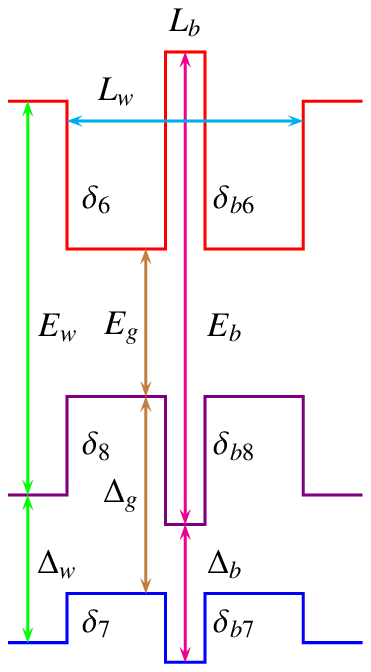}}
  \par\vspace{-0.25cm}
  \caption{(a) Schematic of the band structure of direct gap
    zincblend semiconductors near the $\Gamma$ point ($k=0$).
    The label $\Gamma_{6}$ represents the $s$ states in the conduction band,
    while $\Gamma_{7}$ (split-off holes) and $\Gamma_{8}$ (heavy holes and
    light holes) represent the $p$-states in the valence bands. (b)
    Band offsets for a double quantum well of width $L_{w}$
    with a central barrier of width $L_{b}$. The relationships among
    the several offset parameters are given in
    \protect\eqref{eq:delta87}--\protect\eqref{eq:delta87b}.}
  \label{fig:bands4}
\end{figure}

Following Refs.~\onlinecite{bastard1988} and \onlinecite{winkler2003}
we can straightforwardly generalize the bulk Kane model of
Sec.~\ref{sec:kanebulk} to heterostructures. Essentially, we have
to introduce position-dependent (growth direction) band gaps which
represent the different compounds comprising the heterostructure,
\textit{e.g.}, Fig.~\ref{fig:bands4b}.  In this case, the form of the
resulting Kane Hamiltonian is similar to that of bulk but with
$z$-dependent diagonal matrix elements and with $k_z \rightarrow -i
d/dz$. More specifically, defining $E_6 = H_{11}= H_{22}$,
$E_8=H_{33}= H_{44}=H_{55}=H_{66}$, and $E_7=H_{77}= H_{88}$, we have
for the double quantum well of Fig. 1(b)
\begin{align}
  E_6 &= \frac{\hbar^2 k^{2}}{2m_{0}} + V_{H}(z) +
  h_6(z), \label{eq:E6} \\
  E_8&= \frac{\hbar^2 k^{2}}{2m_{0}} +
  V_{H}(z)- h_8(z) - E_g, \label{eq:E8}\\
  E_7& = \frac{\hbar^2 k^{2}}{2m_{0}} + V_{H}(z) -
  h_7(z) - E_g - \Delta_g, \label{eq:E7}
\end{align}
with $k^2=\kpa^{2}+k_z^2$, $E_g$ and $\Delta_g$ being the
fundamental and split-off band gaps in the well region,
respectively, and
\begin{equation}
\label{eq-offsets}
 h_{i}(z) = \delta_{i}\,h_{w}(z) + \delta_{b\,i}\,h_{b}(z), \qquad
 i=6,7,8,
\end{equation}
where $h_w(z)$ is a dimensionless profile function describing the
shape of a square well of width $L_w$ [and unit depth,
Fig.~\ref{fig:bands4b}]; similarly, $h_b(z)$ describes the shape of
the central square barrier, Fig.~\ref{fig:bands4b}. The parameters
$\delta_6$, $\delta_7$, $\delta_8$, $\delta_{b6}$, $\delta_{b7}$, and
$\delta_{b8}$ denote the relevant band offsets between the well and
the lateral and central barriers for conduction and valence bands.
Defining the zero of energy at the bottom of the conduction well [see
Fig.~\ref{fig:bands4b}], we have
\begin{align}
  \label{eq:delta87}
  \delta_{8} &= E_{w} - E_{g} - \delta_{6}, \quad
  \delta_{7} = \delta_{8} + \Delta_{w} - \Delta_{g}, \\
  \label{eq:delta87b}
  \delta_{b8} &= E_{b} - E_{g} - \delta_{b6}, \quad
  \delta_{b7} = \delta_{b8} + \Delta_{b} - \Delta_{g}.
\end{align}
The corresponding expressions for a single well can readily be
obtained from the above by setting the $\delta_{bi}$'s to zero
(\textit{i.e.}, no central barrier).

Finally, note that we have added a ``Hartree'' potential $V_H(z)$ in
the diagonal elements; see Eqs.~\eqref{eq:E6}--\eqref{eq:E7}. The
Hartree potential $V_H(z)$ here contains contributions from the
electron-electron interaction (mean field) relevant in quantum wells
containing many electrons, the external gate potentials, and the
modulation-doped potential (\textit{i.e.}, ionized impurities outside
the well region). In Appendix~\ref{ap:sc-ap}, we describe in detail
these distinct contributions to $V_H$ and how they are calculated in
our system. As we will see next, both $V_H(z)$ and the structural
confining potentials contribute to the effective SO coupling for
electrons.
\begin{table}[H]
  \caption{\label{tab:basis}  Truncated set of zone center wave functions
   $u_{l\bfs{0}}$ (for simplicity we denote them by $u_{l}$)
   used in constructing the matrix Hamiltonian \eqref{eq-matrix}}
  \begin{ruledtabular}
    \begin{tabular}{cccc}
      $u_{i}$ & $\Gamma$ & \ketd{J}{m_{J}} & $u_{J, m_{J}}$ \\\hline
      $u_{1}$ & $\Gamma_{6}$ & \ketd{\frac{1}{2}}{+\frac{1}{2}} &
        $i\kets{S}{+\frac{1}{2}}$ \\
      $u_{2}$ & $\Gamma_{6}$ & \ketd{\frac{1}{2}}{-\frac{1}{2}} &
        $i\kets{S}{-\frac{1}{2}}$ \\
      $u_{3}$ & $\Gamma_{8}$ & \ketd{\frac{3}{2}}{+\frac{3}{2}} &
        $-\frac{1}{\sqrt{2}}\biggl(\ket{X}+i\ket{Y}\biggr)\otimes
        \ket{+\frac{1}{2}}$ \\
      $u_{4}$ & $\Gamma_{8}$ & \ketd{\frac{3}{2}}{+\frac{1}{2}} &
        $-\frac{1}{\sqrt{6}}\biggl(\ket{X}+i\ket{Y}\biggr)\otimes
        \ket{-\frac{1}{2}} + \sqrt{\frac{2}{3}}\kets{Z}{+\frac{1}{2}}$ \\
      $u_{5}$ & $\Gamma_{8}$ & \ketd{\frac{3}{2}}{-\frac{1}{2}} &
        $+\frac{1}{\sqrt{6}}\biggl(\ket{X}-i\ket{Y}\biggr)\otimes
        \ket{+\frac{1}{2}} + \sqrt{\frac{2}{3}}\kets{Z}{-\frac{1}{2}}$ \\
      $u_{6}$ & $\Gamma_{8}$ & \ketd{\frac{3}{2}}{-\frac{3}{2}} &
        $+\frac{1}{\sqrt{2}}\biggl(\ket{X}-i\ket{Y}\biggr)\otimes
        \ket{-\frac{1}{2}}$ \\
      $u_{7}$ & $\Gamma_{7}$ & \ketd{\frac{1}{2}}{+\frac{1}{2}} &
        $-\frac{1}{\sqrt{3}}\biggl(\ket{X}+i\ket{Y}\biggr)\otimes
        \ket{-\frac{1}{2}} - \frac{1}{\sqrt{3}}\kets{Z}{+\frac{1}{2}}$ \\
      $u_{8}$ & $\Gamma_{7}$ & \ketd{\frac{1}{2}}{-\frac{1}{2}} &
        $-\frac{1}{\sqrt{3}}\biggl(\ket{X}-i\ket{Y}\biggr)\otimes
        \ket{+\frac{1}{2}} + \frac{1}{\sqrt{3}}\kets{Z}{-\frac{1}{2}}$ \\
    \end{tabular}
  \end{ruledtabular}
\end{table}

\section{Effective spin-orbit Hamiltonian for electrons}
\label{sec:so}
\subsection{Folding down}

Since we are interested in SO effects for the conduction electrons,
here we derive an effective Hamiltonian for them. To this end, let us
rewrite our $8\times 8$ Hamiltonian [Eq.~\eqref{eq-matrix}] in the
block form,
\begin{equation}
  \label{eq:H22}
  H_{8\times 8} =
  \begin{pmatrix}
    H_{c} & H_{cv} \\ H_{cv}^{\dagger} & H_{v}
  \end{pmatrix},
\end{equation}
where $H_{c}$ is a $2\times 2$ diagonal matrix in the sector
$\Gamma_{6}$ (conduction band) with identical diagonal elements
$E_{6}$ [\eqref{eq:E6}] and $H_{v}$ is a $6\times 6$ diagonal matrix
in the sectors $\Gamma_{8}$ and $\Gamma_{7}$ (valence bands) with
diagonal elements $E_{8}$ [\eqref{eq:E8}] and $E_{7}$ [\eqref{eq:E7}],
respectively. The $2\times 6$ matrix $H_{cv}$ can be read off directly
from the corresponding $2\times 6$ block in Eq.~\eqref{eq-matrix}.

Using the block form of our Hamiltonian \eqref{eq:H22}) the
eigenvalue problem can be written in the compact form
\begin{equation}
  \label{eq:H22b}
  \begin{pmatrix}
    H_{c} & H_{cv} \\ H_{cv}^{\dagger} & H_{c}
  \end{pmatrix}
  \begin{pmatrix} \varphi_{c} \\ \varphi_{v} \end{pmatrix} =
  E \begin{pmatrix} \varphi_{c} \\ \varphi_{v} \end{pmatrix},
\end{equation}
where $\varphi_{c}$ is a two-component spinor (conduction sector)
and $\varphi_{v}$ is a six-component spinor (valence sector).
Straightforward manipulations\cite{bastard1988,winkler2003} yield
the effective Schr{\"{o}}dinger's equation
\begin{equation}
  \mathcal{H}(E) \tilde{\varphi}_{c}  = E \tilde{\varphi}_{c},
\label{Sch_eff}
\end{equation}
with
\begin{equation}
  \mathcal{H}(E)
  =H_{c}+H_{cv}\bigl(E-H_{v}\bigr)^{-1}H_{cv}^{\dagger}, \label{eq:Heff}
\end{equation}
and $\tilde{\varphi}_{c}$ is a properly renormalized
conduction-electron spinor.\cite{renorm}

The matrix elements of $\mathcal{H}(E)$ are given by\cite{neglect-terms}
\begin{align}
  {\mathcal{H}}(E)_{11} &= {\mathcal{H}}(E)_{22} = E_{6} +
  \frac{P^{2}}{3}\bigl(\kpa^{2}\,\gamma_{1} + k_{z}\,\gamma_{1}k_{z}\bigr),
  \label{eq:Hl11}\\
  {\mathcal{H}}(E)_{12} &= {\mathcal{H}}(E)_{21}^{\dagger} =
  \frac{P^{2}}{3}k_{-} \bigl[\gamma_{2},k_{z}\bigr] = -\frac{P^{2}}{3}
  k_{-}k_{z}\,\gamma_{2}, \label{eq:Hl12}
\end{align}
where $\kpa^{2}=k_{\pm}k_{\mp}=k_{x}^{2}+k_{y}^{2}$ and
\begin{align}
  \gamma_{1}(z) &= \biggl(\frac{2}{E - E_{8}} +
  \frac{1}{E - E_{7}}\biggr), \label{eq:gama1} \\
  \gamma_{2}(z) &= \biggl(\frac{1}{E - E_{8}} -
  \frac{1}{E - E_{7}}\biggr), \label{eq:gama2}
\end{align}
We should emphasize that Eq.~\eqref{Sch_eff} is not really an
eigenvalue equation as $\mathcal{H}(E)$ depends on $E$. However, as we
show in Sec.~\ref{subsec:expans}, we can still obtain a true
eigenvalue problem by performing suitable expansions.

\subsection{Energy denominator expansions}
\label{subsec:expans}

Since $E_{g}$ and $E_{g}+\Delta$ are the largest energy scales in
our system, \textit{i.e.},
\begin{align}
 \chi_{8} &= \frac{E-\left[\dfrac{\hbar^2 k^{2}}{2m_{0}} + V_{H}(z)-
  h_8(z)\right]}{E_{g}}\ll 1, \label{eq:xi8}\\
  \chi_{7} &= \frac{E-\left[\dfrac{\hbar^2 k^{2}}{2m_{0}} + V_{H}(z)-
  h_7(z)\right]}{E_{g}+\Delta_{g}}\ll 1,
  \label{eq:xi7}
\end{align}
we can expand the energy denominators in the $\gamma_{i}$'s
[Eqs.~\eqref{eq:gama1} and \eqref{eq:gama2}] in the form
\begin{align}
  \gamma_{1} &= \frac{2}{E_{g}}\bigl(1-\chi_{8}+\cdots\bigr) +
  \frac{1}{E_{g}+\Delta_{g}}\bigl(1-\chi_{7}+\cdots\bigr) \label{eq:a1t} \\
  \gamma_{2} &= \frac{1}{E_{g}}\bigl(1-\chi_{8}+\cdots\bigr) -
  \frac{1}{E_{g}+\Delta_{g}}\bigl(1-\chi_{7}+\cdots\bigr)\label{eq:a2t}.
\end{align}
For the diagonal matrix elements ${\mathcal{H}}(E)_{11} =
{\mathcal{H}}(E)_{22}$ we keep only zeroth-order (\textit{i.e.},
energy-independent) terms, while for the off-diagonal matrix
elements ${\mathcal{H}}(E)_{12} = {\mathcal{H}}(E)_{21}^{\dagger}$
we keep in addition the first-order terms as they give the lowest
non-vanishing contribution (because the off-diagonal matrix elements
contain derivatives with respect to $z$). Straightforwardly, we then
obtain the \textit{energy-independent} one electron Hamiltonian
\begin{equation}
  \label{eq:Hc}
  {\mathcal{H}}(E) =
  H_{QW}\mathbf{1} + \eta(z)\,
  \begin{pmatrix}
    0 & -i k_{-} \\
    i k_{+} & 0
  \end{pmatrix},
\end{equation}
where
\begin{equation}
  \label{eq:QWH}
  H_{QW}=\frac{\hbar^2\kpa^{2}}{2\eem} +
  \frac{\hbar^2}{2\eem}\frac{\partial^2}{\partial z^2}+V_{sc}(z),
\end{equation}
[the subscript ``\textit{sc}'' emphasizes that the potential is to be
determined self-consistently -- see Appendices \ref{ap:sc-ap} and
\ref{apx:cis}] and\cite{prl99eb2007}
\begin{align}
  \label{eq:Mtrans0}
  \frac{1}{\eem} &= \frac{1}{m_{0}} +
  \frac{2P^{2}}{3\hbar^{2}}\biggl(\frac{2}{E_{g}} +
  \frac{1}{E_{g}+\Delta_{g}}\biggr), \\
  \label{eq:Vtot}
  V_{sc}(z) &= V_{H}(z) + \delta_{6}h_w(z) + \delta_{b6}h_{b}(z), \\
  \label{eq:eta1}
  \eta(z) &= \eta_{w}\frac{dh_w(z)}{dz} +
  \eta_{b}\frac{dh_{b}(z)}{dz} - \eta_{H}\frac{dV_{H}(z)}{dz},
\end{align}
with
\begin{align}
  \eta_{H} &= \frac{P^{2}}{3} \biggl[ \frac{1}{E_{g}^{2}}
  - \frac{1}{(E_{g}+\Delta_{g})^{2}} \biggr], \label{eq:etaH} \\
  \eta_{w} &= \frac{P^{2}}{3} \biggl[
  \frac{\delta_{8}}{E_{g}^{2}} -
  \frac{\delta_{7}}{(E_{g}+\Delta_{g})^{2}} \biggr], \label{eq:etaw} \\
  \eta_{b} &= \frac{P^{2}}{3} \biggl[
  \frac{\delta_{b8}}{E_{g}^{2}} -
  \frac{\delta_{b7}}{(E_{g}+\Delta_{g})^{2}} \biggr]. \label{eq:etab}
\end{align}

\subsection{Projection into the quantum well subbands}
\label{sec:proj}

Here we define a quasi-two-dimensional (2D) model starting from the
three-dimensional (3D) Hamiltonian \eqref{eq:Hc}. The idea is
essentially to obtain a 2D effective model similar to the well-known
Rashba model, but now for the case of wells with many subbands. To
this end we (i) first project [\eqref{eq:Hc} into the spin-degenerate
eigenstates of $H_{QW}$ [Eq.~\eqref{eq:QWH}] (note that $H_{QW}$
\textit{does not} contain the SO interaction):\cite{zawadzki-review}
$|\kpab v\rangle_{\sigma _{z}}= |\kpab v\rangle |\sigma_{z}\rangle$,
$\langle\bfl{r}|\kpab v\rangle =\exp(i\kpab\cdot\rpab)\varphi_{v}(z)$,
$v=0,1,2,\ldots$, and $\sigma_{z}=\pm$ (or $\uparrow, \downarrow$),
which correspond to the subband energies $\mce_{\kpa v} =
\frac{\hbar^2\kpa^{2}}{2\eem} +\mce_{v}$, with $\mce_{v}$ being the
quantized levels of the well, and then (ii) consider a reduced set of
subbands (\textit{e.g.}, two) by truncating the basis set used. In this
section we simply assume that we know the eigensolutions of $H_{QW}$;
later on we actually calculate them within a self-consistent
procedure, from which we can explicitly determine the relevant SO
coupling constants in our problem.

The matrix elements of ${\mathcal{H(\varepsilon)}}$ in the
$\{|\kpab v\rangle_{\sigma_{z}}\}$ basis are
\begin{align}
  \label{eq:EH1}
  \bra{\kpab v} \bra{\pm} {\mathcal{H}(E)}
  \ket{\kpab v'} \ket{\pm} &= \bigl(\frac{\hbar^2\kpa^{2}}{2\eem} +
  \mce_{v}\bigr)\, \delta_{vv'}, \\
  \label{eq:EH2}
  \bra{\kpab v} \bra{\pm} {\mathcal{H}(E)}
  \ket{\kpab v'} \ket{\mp} &= \mp i\,\eta_{vv'}\,k_{\mp},
\end{align}
with the generalized SO couplings
\begin{equation}
  \label{eq:etatot}
  \eta_{vv'} = \Gamma_{vv'}^{H} + \Gamma_{vv'}^{w} +
  \Gamma_{vv'}^{b},
\end{equation}
where
\begin{align}
  \label{eq:EtaH}
  \Gamma_{vv'}^{H} &= -\eta_{H}\bra{v}\frac{dV_{H}(z)}{dz}\ket{v'}, \\
  \label{eq:EtaW}
  \Gamma_{vv'}^{w} &= +\eta_{w}\bra{v}\frac{dh_w(z)}{dz}\ket{v'}, \\
  \label{eq:EtaB}
  \Gamma_{vv'}^{b} &= +\eta_{b}\bra{v}\frac{dh_{b}(z)}{dz}\ket{v'}.
\end{align}
The coefficients $\Gamma_{vv'}^{H}$, $\Gamma_{vv'}^{w}$, and
$\Gamma_{vv'}^{b}$ denote the contributions from the Hartree
potential, the quantum-well edges, and the central barrier edges,
respectively. It is convenient to split the Hartree contribution into
two terms, \textit{i.e.}, $V_H(z)=V_e(z)+V_g(z)$, where $V_e(z)$ is
the purely electronic Hartree potential and $V_g(z)$ denotes the
contributions from the external gate potential and the modulation
doping potential. Hence $\Gamma_{vv'}^{(H)} =
-\eta_{H}\bra{v}\frac{dV_{e}(z)}{dz}\ket{v'}
-\eta_{H}\bra{v}\frac{dV_{g}(z)}{dz}\ket{v'}$. This separation will be
useful when discussing our results.

We emphasize that the diagonal (in $v,v'$) parameters $\eta_{vv}$
correspond to the Rashba coupling in the $v$th subband, \textit{i.e.},
$\alpha_v = \eta_{vv}$. The off-diagonal terms $\eta_{vv'}$ arise due
to the intersubband coupling. Interestingly, these new SO terms can be
non-zero even in structurally symmetric wells, since they arise from
quantum-well states of distinctive parities.

For completeness we present here the linearized Dresselhaus
couplings\cite{irene} in the $v$th subband
\begin{equation}
  \label{eq:Dress}
  \beta_{v} = \beta_{D}\bra{v}k_{z}^{2}\ket{v},
\end{equation}
where the constant $\beta_{D}$ is the bulk Dresselhaus SO parameter.
\cite{prb72jmj2005} We can easily rewrite the above expression in
the more convenient form
\begin{equation}
  \label{eq:Dress2}
  \beta_{v} = \beta_{D}\frac{2\eem}{\hbar^{2}}
  \left[\mce_{v} - \bra{v} V(z) \ket{v} \right].
\end{equation}
In Sec.~\ref{sec:results} we shall use the above form to discuss how
the Dresselhaus couplings vary as a function of the system parameters.

\subsection{Two-subband case}
To illustrate the procedure of Sec.~\ref{sec:proj}, let us explicitly
work out here the case of a quantum well with only two subbands
$v=0,1$. In Sec.~\ref{sec:results} we shall investigate in detail the
SO couplings for single and double quantum wells with two subbands.

\subsubsection{4x4 Hamiltonian}
With the basis ordering $\{|\kpab 0 \rangle_{\uparrow}, |\kpab 0
\rangle_{\downarrow}, |\kpab 1 \rangle_{\uparrow}, |\kpab 1
\rangle_{\downarrow}\}$, Eqs.~\eqref{eq:EH1} and \eqref{eq:EH2} yield
the effective Hamiltonian
\begin{equation}
  \label{eq:EH}
  H =
  \begin{pmatrix}
    \mce_{\kpa 0} & -i\alpha_{0}\, k_{-} & 0 & -i\eta\, k_{-} \\
    i\alpha_{0}\, k_{+} & \mce_{\kpa 0} & i\eta\, k_{+} & 0 \\
    0 & -i\eta\, k_{-} & \mce_{\kpa 1} & -i\alpha_{1}\, k_{-} \\
    i\eta\, k_{+} & 0 & i\alpha_{1}\, k_{+} & \mce_{\kpa 1}
  \end{pmatrix},
\end{equation}
where the Rashba couplings are given by $\alpha_{v}=\eta_{vv}$,
$v=0,1$, and the intersubband SO coupling\cite{referee} by
$\eta=\eta_{01}$ [see Eqs.~\eqref{eq:etatot} and \eqref{eq:EtaB}] and
\begin{equation}
  \label{eq:ekv}
  \mce_{\kpa v} = \mce_{v} + \frac{(\hbar \kpa)^{2}}{2\eem},\quad v=0,1.
\end{equation}

\subsubsection{Eigensolutions}
The energy eigenvalues $\mce_{\sigma\lambda}$ of Eq.~\eqref{eq:EH} are
straightforward to obtain:
\begin{equation}
  \label{eq:av}
  \mce_{\kpa,\lambda_{1},\lambda_{2}} = \mce_{\kpa +} +
  \lambda_{2}\alpha_{+}\,\kpa + \lambda_{1} \sqrt{(\eta\kpa)^{2} +
  (\mce_{\kpa -} + \lambda_{2}\alpha_{-}\kpa)^{2}},
\end{equation}
where $\lambda_{2}=\pm$ are spin quantum numbers and
$\lambda_{1}=\pm$ are the subband (or pseudo spin) indices, and
\begin{equation}
  \label{eq:aux2}
  \mce_{\kpa \pm} = \frac{1}{2}\big(\mce_{\kpa 1}\pm\mce_{\kpa 0}\big),\quad
  \alpha_{\pm} = \frac{1}{2}\big(\alpha_{1}\pm\alpha_{0}).
\end{equation}

The corresponding (normalized) eigenvectors are
\begin{equation}
  \ket{\lambda_{1},\lambda_{2}} =
  \sqrt{1+\lambda_{1}\frac{\ep_{\lambda_{2}}(0)}
  {\ep_{\lambda_{2}}(\eta)}}
  {\renewcommand{\arraystretch}{2.0}
  \begin{pmatrix}
    \cfrac{-i\lambda_{1}\eta \kpa\,\e^{-i\theta}}
    {\ep_{\lambda_{2}}(\eta)+\lambda_{1}\ep_{\lambda_{2}}(0)} \\
    \cfrac{\lambda_{1}\lambda_{2}\eta \kpa}
    {\ep_{\lambda_{2}}(\eta)+\lambda_{1}\ep_{\lambda_{2}}(0)} \\
    -i\lambda_{2}\e^{-i\theta} \\
    1
  \end{pmatrix}}\,\dfrac{\e^{i\,\kpab\cdot\rpab}}{4\pi},
\end{equation}
where
\begin{equation}
  \label{eq:aux1}
  \ep_{\pm}(\eta) = \sqrt{(\eta\kpa)^{2} + (\mce_{\kpa -}\pm
  \alpha_{-}\kpa)^{2}},\quad \e^{\pm i\theta} = \frac{k_{\pm}}{\kpa}.
\end{equation}

\subsubsection{SO-induced effective mass renormalization}

Expanding the energy dispersions [Eq.~\eqref{eq:av}] around $\kpa=0$,
we obtain to second order
\begin{equation}
  \label{eq:RDk0}
  \mce_{\kpa\to 0,\lambda_{1},\lambda_{2}}\approx \mce_{+} +
  \lambda_{1}\,\mce_{-} +
  \lambda_{2}\,(\alpha_{+}+\lambda_{1}\,\alpha_{-})\kpa +
  \frac{\hbar^2\kpa^{2}}{2\eem_{\lambda_{1}}},
\end{equation}
where $\eem_{\lambda_{1}}$ are the effective masses
\begin{equation}
  \label{eq:SOEM}
  \eem_{\pm} = \cfrac{\eem}{1\pm \cfrac{2\mce_{so}}{\Delta\mce}},
\end{equation}
where $\mce_{so}=\frac{1}{2}\eem\eta^2/\hbar^2$ and
$\Delta\mce=2\mce_{-}$. Note that the mass renormalization is solely
due to the intersubband-induced SO coupling $\eta$. For the
realistic wells we investigate here  $2\mce_{so}/\Delta\mce<<1$ for
single wells but can reach $\sim 0.1$ for double wells (Secs.
\ref{sec:systems} and \ref{sec:results}).

\subsubsection{Determining the SO couplings}

As mentioned previously, we determine the SO orbit couplings (here
specifically $\alpha_0$, $\alpha_1$, and $\eta$) from the
self-consistent eigensolutions of the quantum well \textit{without}
spin orbit,\cite{zawadzki-review} via
Eqs.~\eqref{eq:etatot})-\eqref{eq:EtaB}). In Sec.~\ref{sec:systems} we
detail the quantum well systems investigated and briefly outline the
self-consistent procedure used to obtain the eigensolutions (a full
description is provided in Appendices~\ref{ap:sc-ap} and
\ref{apx:cis}). We then present results for single and double wells
with two subbands; \textit{i.e.}, we calculate $\alpha_0$, $\alpha_1$,
and $\eta$ and discuss in detail the several distinct contributions to
each of these quantities.

\section{Quantum-well systems and self-consistency}
\label{sec:systems}
\begin{figure}[h]
  \centering
  \includegraphics{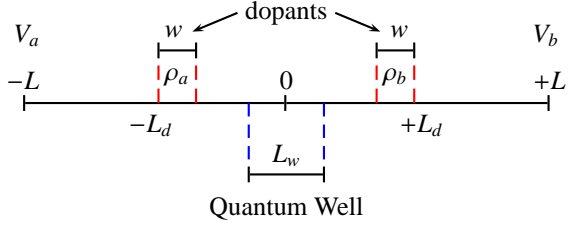}
  \caption{Schematic view of our quantum-well system. The doping
    densities $\protect\rho_{a}$, $\protect\rho_{b}$ and the external
    gate voltages $V_{a}$ and $V_{b}$ can be used to control the
    degree of the structural inversion asymmetry.}
  \label{fig:qwell}
\end{figure}

Figure~\ref{fig:qwell} shows a schematic view of the quantum-well
system we study: a well of width $L_{w}$ centered at $z=0$ ($z$:
growth direction), and two adjacent symmetrically doped regions of
widths $w$ in the barriers. We also consider double wells by inserting
an additional (central) barrier of width $L_b$ in the well region. The
doping densities of the left and right regions, $\rho_{a} $ and
$\rho_{b}$, respectively, can be used to control the degree of
structural inversion asymmetry of the wells (in
Sec.~\ref{sec:results}, however, we present results only for
$\rho_{a}=\rho_{b}$). The external gates $V_{a}$ and $V_{b}$, located
at the end points $\pm L$, can also be used to control the degree of
inversion asymmetry and to vary the areal electron density in the
well.

Since our wells have many electrons and are subject to external gates,
we have to solve the Schr{\"{o}}dinger and Poisson equations
self-consistently (``Hartree approximation''\cite{enderlein1999}) in
order to determine their potential profile $V_{sc}(z)$ [see
Eq.~\eqref{eq:Vtot}] and corresponding eigenfunctions and
eigenenergies.  In Appendices~\ref{ap:sc-ap} and \ref{apx:cis} we
describe in detail our standard self-consistent procedure.

\begin{figure}[h]
  \centering
  \subfigure[\label{fig:GaInAsPot3a} Self-consistent potential energies.]
  {\resizebox{\imsizeB}{!}{\includegraphics{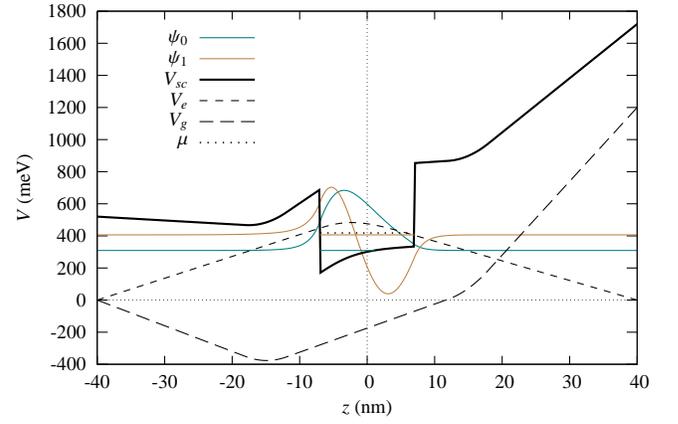}}}
  \subfigure[\label{fig:GaInAsPot3b} Force fields.]
  {\resizebox{\imsizeF}{!}{\includegraphics{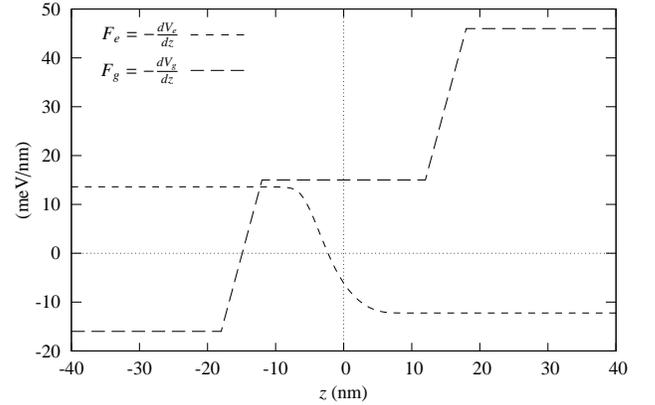}}}
  \par\vspace{-0.25cm}
  \caption{(a): Self-consistent potential energy $V_{sc}$ (thick solid
    line) and the corresponding wave functions $\psi_{0}$ and
    $\psi_{1}$ for the single well
    Al$_{0.48}$In$_{0.52}$As/Ga$_{0.47}$In$_{0.53}$As with external
    gates $V_{a}=0$ e $V_{b}=1.2$~eV (see
    Fig.~\protect\ref{fig:qwell}). The the electronic Hartree
    potential $V_{e}$ (short dashed line), the external gate plus
    modulation doping contributions $V_{g}$ (long dashed line), and
    the corresponding force fields $F_{e}=-dV_e/dz$ and
    $F_{g}=-dV_g/dz$ are also shown in (b). The two levels in the well
    (solid lines) denote the energies of the first and second subband
    edges, while the dotted level indicates the chemical potential.}
\end{figure}

Before going into the discussion of the SO couplings in detail, let
us first have a look at the outcome of a typical self-consistent
simulation we perform. Figure~\ref{fig:GaInAsPot3a} shows the
self-consistent potential $V_{sc}$ (thick solid line) for a single
well with two subbands; the corresponding self-consistent wave
functions $\psi_0(z)$ and $\psi_1(z)$ are also shown. The energies
of the two lowest subband edges (see levels in the well) are
$\mathcal{E}_{0}=309.09$~meV and $\mathcal{E}_{1}=406.39$~meV ($
\Delta\mathcal{E} = 97.3$~meV). Here we fix the chemical potential
at $\mu=413.40$~meV with respect to the $V=0$ origin (``constant
chemical potential model'', see below) and set the external gates to
$V_{a}=0$ and $V_{b}=1200$~meV. The two subbands are occupied with
areal densities $n_{0}=18.7422\times 10^{11}$~cm$^{2}$ and
$n_{1}=1.2578\times 10^{11}$~cm$^{2}$, respectively. The electronic
Hartree potential $V_e$ (short dashed line) and the the external
gates (plus modulation doping) contribution $V_g$ (long dashed line)
are also shown. Figure \ref{fig:GaInAsPot3b} shows the corresponding
``force fields'' $F_{e}=-dV_{e}/dz$ arising from the confined
electrons in the well and $F_{g}=-dV_{g}/dz$ coming from the doping
regions ($\pm 12$~nm to $\pm 18$~nm) plus the external gates ($F_g$
and $F_e$ will be useful when discussing the SO couplings further
below). Using the self-consistent solutions $\psi_v(z)$, $v=0,1$, we
can straightforwardly calculate the relevant SO couplings [via Eqs.
\eqref{eq:etatot}--\eqref{eq:EtaB}]: $\eta=-3.81$~meV\,nm,
$\alpha_{0}=-5.44$~meV\,nm, $ \alpha_{1}=-3.74$~meV\,nm,
$\beta_{0}=0.87$~meV\,nm, and $\beta_{1}=2.50$~meV\,nm.

\section{Results}
\label{sec:results}
Here we focus on single and double quantum wells with only two
subbands. More specifically, we calculate three SO couplings: the
intersubband-induced SO coupling $\eta=\eta_{01}$ and the two
Rashba-type couplings $\alpha_{0}=\eta_{00}$ and
$\alpha_{1}=\eta_{11}$. We consider two experimentally relevant cases:
the constant areal density ($n_T$-constant) and the constant chemical
potential ($\mu$-constant) models. In our simulations we always keep
$V_{a}=0$ as a reference potential and vary $V_{b}$; see
Fig.~\ref{fig:qwell}. For completeness, we also calculate the two
Dresselhaus constants $\beta_0$ and $\beta_1$ [see
Eq.~\eqref{eq:Dress}] within each subband.

\subsection{Single wells}
\label{sec:GaInAs}

\begin{table}[h]
  \caption{Relevant parameters\cite{jap89iv2001} (see Fig.~\ref{fig:bands4})
    (at 0.3~K) for the single quantum well
    Al$_{0.48}$ In$_{0.52}$As/Ga$_{0.47}$In$_{0.53}$As system in our
    simulations. The doping regions have widths $ w=6$~nm and densities
    $\protect\rho_{a}=\protect\rho_{b}=4\times 10^{18}$~cm$^{-3}$ (see
    Fig.~\ref{fig:qwell}). All energies are in eV and lengths in nm;
    the coefficient $\eta_{H}$ is in nm$^{2}$ and $\eta_{w}$ in
    meV\,nm$^{2}$. The Dresselhaus coupling
    constant\cite{prb72jmj2005} $\beta{D}$ is in meV\,nm$^{3}$.}
  \label{tab:bpGaInAs}
  \begin{ruledtabular}
    \begin{tabular}{rcl|rcl|rcl|rcl}
      $E_{w}$&=&$1.5296$ & $\Delta_{w}$&=&$0.2998$ & $w$ &=& $6$ &
      $E_{P}$&=&$25.3$ \\ \hline
      $E_{g}$&=&$0.8161$ & $\Delta_{g}$&=&$0.3296$ & $\delta_{6}$&=&$0.52$ &
      $\eem/m_{0}$&=&$0.043$ \\ \hline
      $E_{b}$&=&$0$ & $\Delta_{b}$&=&$0$ & $\delta_{b6}$&=&$0$ &
      $\epsilon_{r}$&=&$14.013$ \\ \hline
      $L$&=&$40$ & $L_{d}$&=&$18$ & $L_{b}$&=&$0$ & $L_{w}$&=&$14$ \\ \hline
      $\eta_{H}$&=&$0.2376$ & $\eta_{w}$&=&$0.0533$ & $\eta_{b}$&=&$0$ &
      $\beta{D}$&=&$0.0237$\\
    \end{tabular}
  \end{ruledtabular}
\end{table}

\subsubsection{Single-well parameters}

We consider a realistic
Al$_{0.48}$In$_{0.52}$As/Ga$_{0.47}$In$_{0.53}$As single quantum
well.\cite{prl89tk2002,prb74tk2006} We assume doping densities
$\rho_{a}=\rho_{b}=4\times 10^{18}$~cm$^{-3}$ with widths $w=6$~nm
(``sample 3'' in Ref.~\onlinecite{prl89tk2002}).
Table~\ref{tab:bpGaInAs} summarizes band parameters, potential
offsets,\cite{jap89iv2001} well widths, and other important parameters
of our system. The coefficients $\eta_{w}$ and $\eta_{H}$ in
Tab.~\ref{tab:bpGaInAs} are defined in Eq.~\eqref{eq:etaw} and
Eq.~\eqref{eq:etaH}, respectively. Here, the Dresselhaus parameter
$\beta{D}$ in Eq.~\eqref{eq:Dress} is assumed to be the same as that
of the GaAs (see Ref.~\onlinecite{prb72jmj2005}).

\subsubsection{SO couplings: single wells}

\begin{figure}[h]
  \centering
  \subfigure[\label{fig:GaInAsCC1} $n_{T}$ constant.]
    {\resizebox{\imsizeC}{!}{\includegraphics{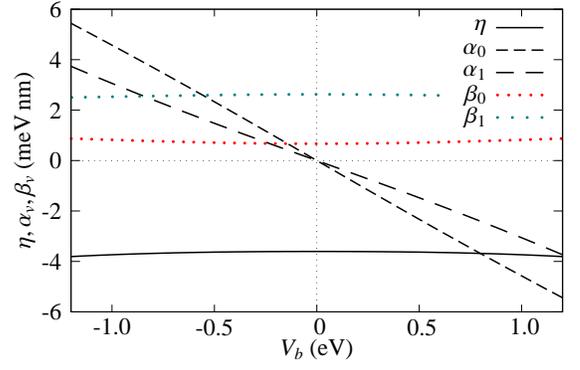}}}\\
  \subfigure[\label{fig:GaInAsCC2} $\mu$ constant.]
    {\resizebox{\imsizeC}{!}{\includegraphics{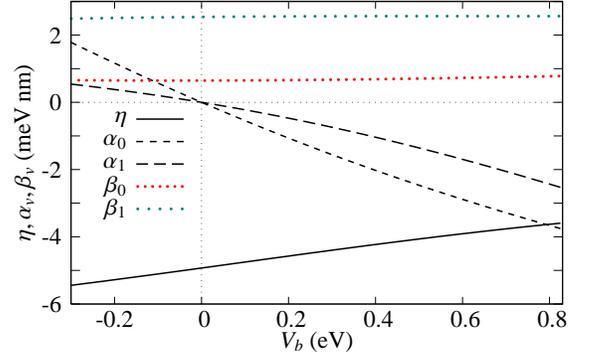}}}
  \par\vspace{-0.25cm}
  \caption{Rashba $\alpha$, Dresselhaus $\beta$ and intersubband-induced
    $\eta$ SO coupling constants for the
    Al$_{0.48}$In$_{0.52}$As/Ga$_{0.47}$In$_{0.53}$As quantum well as
    functions of the gate voltage $V_{b}$ (see Fig.~\ref{fig:qwell}).
    In (a) the total 2D electron density is kept constant at
    $n_{T}=20\times 10^{11}$~cm$^{-2}$ and in (b) the chemical
    potential is kept constant at $\mu=200$~meV.}
  \label{fig:GaInAsCC}
\end{figure}

Figure~\ref{fig:GaInAsCC} shows the strength of the Rashba
($\alpha_\nu$, $ \nu=0,1$; dashed lines), Dresselhaus ($\beta_\nu$,
$\nu=0,1$; dotted lines) and intersubband-induced ($\eta$, solid
line) SO couplings as functions of the gate voltage $V_{b}$, for
both the $n_T$-constant and the $\mu$-constant models,
Figs.~\ref{fig:GaInAsCC1} and \ref{fig:GaInAsCC2}, respectively. At
$V_{b}=V_{a}=0$~eV, our sample is completely symmetric and, as
expected, the Rashba couplings $\alpha_0$ and $\alpha_1$ are zero.
We note that the Dresselhaus couplings $\beta_0$ and $\beta_1$ are
practically constant in both models. This follows from Eq.
\eqref{eq:Dress2} which shows that in each subband the Dresselhaus
coupling is essentially the difference between the expected value of
the self-consistent potential in the respective subband and the
corresponding eigenenergy. The Rashba couplings, on the other hand,
vary considerably with $V_b$, although showing a similar trend in
both models. Interestingly, they change signs about $V_b=0$
(symmetric configuration), but always with
$|\alpha_{0}|>|\alpha_{1}|$. Our calculated $\alpha_0$ within the
$\mu$-constant model [Fig.~\ref{fig:GaInAsCC2}] is consistent with
the measurements of this quantity by Koga \textit{et
al.},\cite{prl89tk2002,prb74tk2006} whose samples have a constant
chemical potential.

The new intersubband-induced coupling $\eta$ [see the solid lines in
Figs.~\ref{fig:GaInAsCC1} and ~\ref{fig:GaInAsCC2}] is non-zero even
in the symmetric well configuration ($V_{b}=0=V_a$). It has a strength
comparable to the Rashba and is at least twice as large as the
Dresselhaus. In contrast to the Rashba couplings, the intersubband SO
$\eta$ does not change sign with $V_b$. In fact, for the single well
investigated here $\eta$ is almost constant with $V_b$, although it
varies slightly more in the $\mu$-constant model [compare the solid
curves in Figs.~\ref{fig:GaInAsCC1} and \ref{fig:GaInAsCC2}].

\begin{figure}[H]
  \centering
   \subfigure[\label{fig:GaInAsE1a} Contributions to $\eta$.]
    {\resizebox{\imsizeC}{!}{\includegraphics{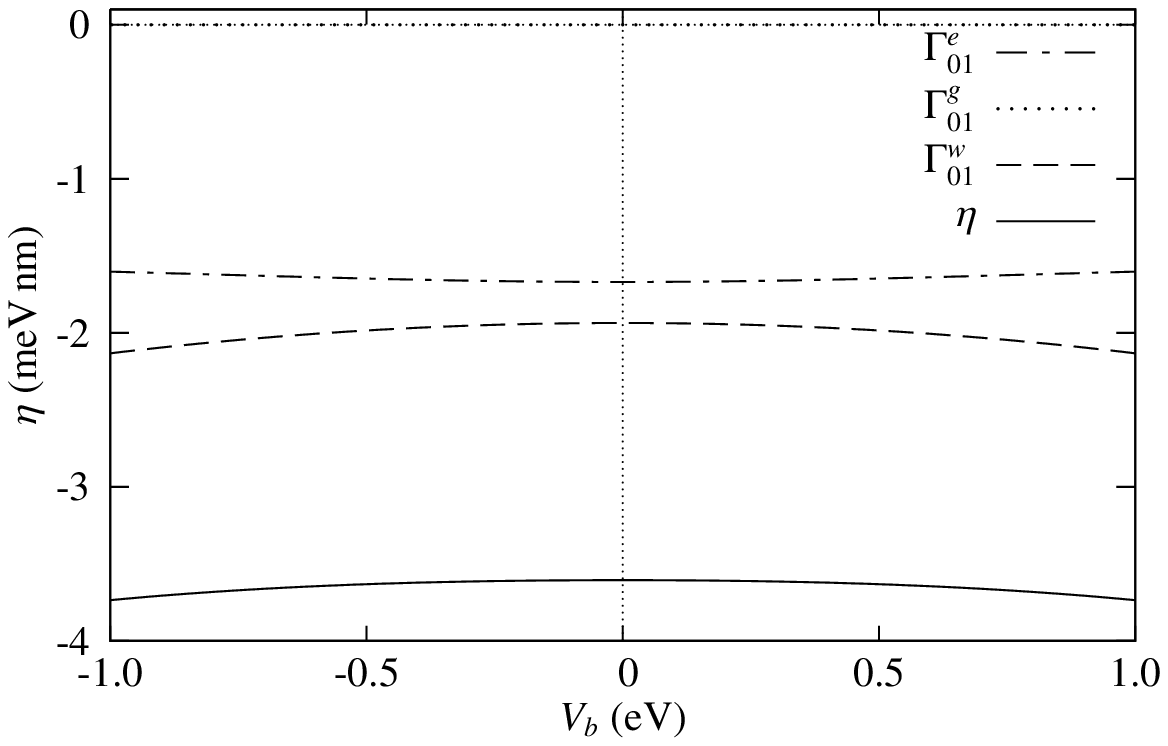}}}\\
  \subfigure[\label{fig:GaInAsA0a} Contributions to $\alpha_{0}$.]
    {\resizebox{\imsizeC}{!}{\includegraphics{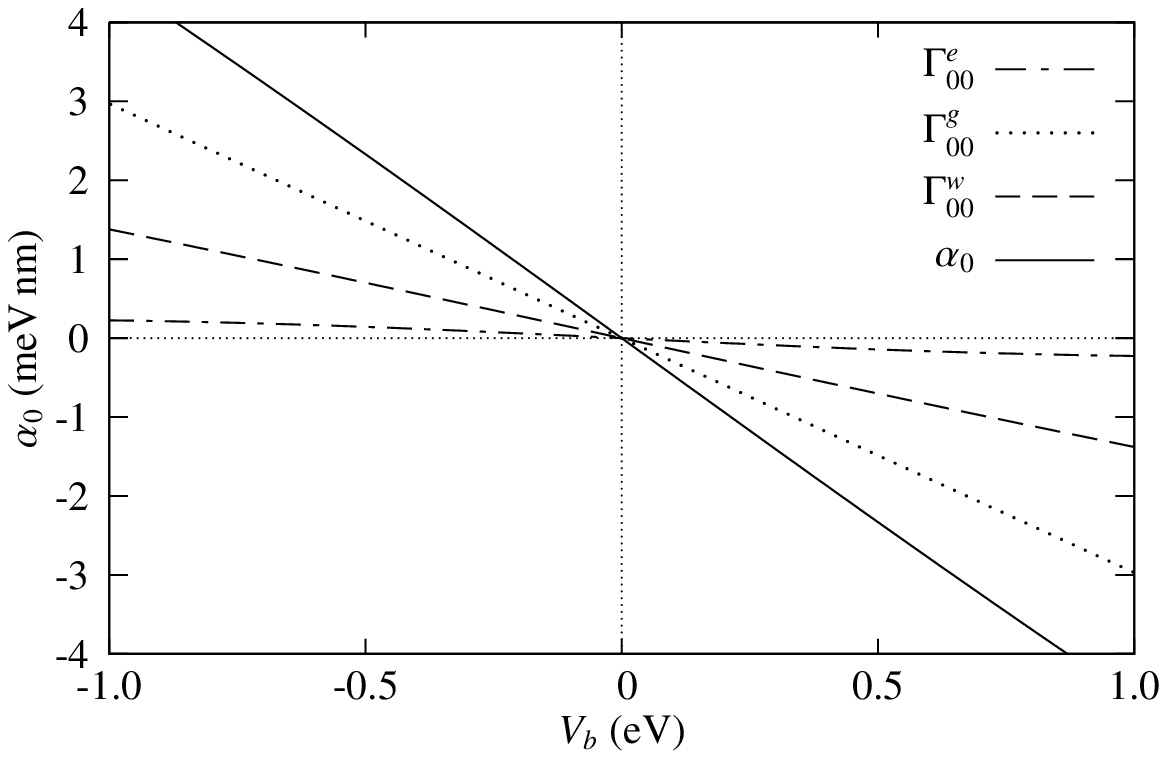}}}\\
  \subfigure[\label{fig:GaInAsA1a} Contributions to $\alpha_{1}$.]
    {\resizebox{\imsizeC}{!}{\includegraphics{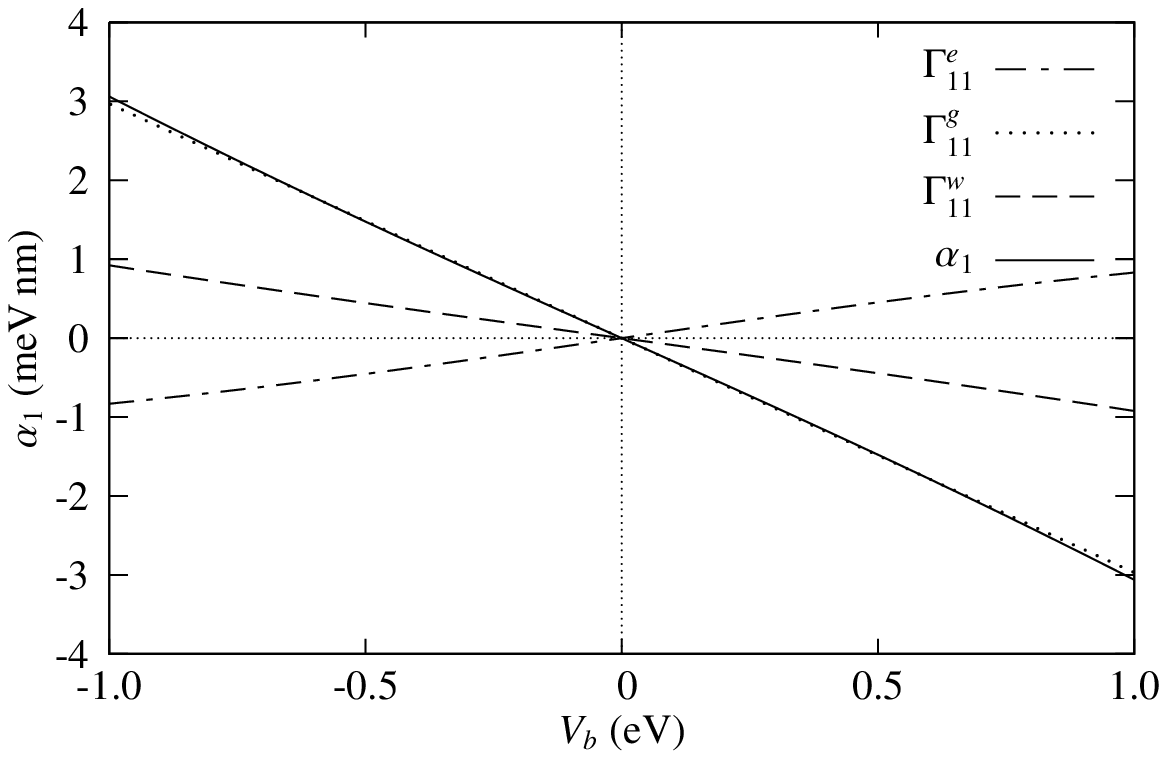}}}
  \par\vspace{-0.25cm}
  \caption{\label{fig:GaInAsAE} Several distinct contributions to the
    coupling constants $\eta$ (a), $\alpha_{0}$ (b) and $\alpha_{1}$
    (c) for the single GaInAs quantum well shown in
    Fig.~\ref{fig:GaInAsCC1} ($n_T$-constant model) as functions of the
    external gate $V_{b}$ ($V_{a}=0$). These contributions arise
    from: the electron density (Hartree potential), the external gate
    (together with donor regions), and the structural well potential;
    these are denoted by the superscripts e, g and w,
    respectively.}
\end{figure}

To more easily understand the results above, we analyze the several
contributions to the SO couplings separately. To this end, we
rewrite [see comments following Eq.~\eqref{eq:etatot}] $\eta_{vv'}$
for a single well in the form
\begin{equation}
\label{eq:eta-sqw} \eta_{vv'}^{SW} = \Gamma_{vv'}^{e} +
\Gamma_{vv'}^{g} + \Gamma_{vv'}^{w},
\end{equation}
where we have set $\Gamma_{vv'}^{b}=0$ in \eqref{eq:etatot},
\textit{i.e.}, no central barrier contribution, and have split the
Hartree contribution into its purely electronic $\Gamma_{vv'}^{e}$ and
the external gate (plus doping potential) $\Gamma_{vv'}^{g}$ parts.
Hence, for two subbands, each of the SO couplings has three
contributions: $\eta=\eta_{01}^{SW}= \Gamma_{01}^{e}+
\Gamma_{01}^{g}+\Gamma_{01}^{w}$, $\alpha_0=\eta_{00}^{SW}=
\Gamma_{00}^{e}+ \Gamma_{00}^{g}+\Gamma_{00}^{w}$ and
$\alpha_1=\eta_{11}^{SW} = \Gamma_{11}^{e}+
\Gamma_{11}^{g}+\Gamma_{11}^{w}$.
Figures~\ref{fig:GaInAsE1a}--\ref{fig:GaInAsA1a} show the above
contributions separately for the $n_T$-constant case (similar results
hold for the $\mu$-constant model, in the parameter range studied).

Figure~\ref{fig:GaInAsE1a} shows that the external gates and doping
contributions to $\eta$ ($\Gamma_{01}^{g}$ curve) are essentially
zero, while the electronic Hartree contribution ($\Gamma_{01}^{e}$
curve) and the structural ($\Gamma_{01}^{w}$ curve) contributions are
comparable in magnitudes and both negative. In contrast, for both
$\alpha_0$ and $\alpha_1$ the largest contributions come from the
external gates together with doping regions [see the curve
$\Gamma_{00}^{g}$ in Fig.~\ref{fig:GaInAsA0a} and the curve
$\Gamma_{11}^{g}$ in Fig.~\ref{fig:GaInAsA1a}]; these account for 60\%
of $\alpha_0$ and 100\% of $\alpha_1$. The electronic Hartree
contribution is negligible in $\alpha_0$ [curve $\Gamma_{00}^{e}$ in
Fig.~\ref{fig:GaInAsA1a}] while the structural part
($\Gamma_{00}^{w}$) accounts for about 30\% of it. On the other hand,
the structural and electronic Hartree contributions in $\alpha_1$
essentially cancel out (same magnitude and opposite signs);
\textit{cf.} the $\Gamma_{11}^{e}$ and $\Gamma_{11}^{w}$ curves in
Fig.~\ref{fig:GaInAsA1a}.

We can understand the above remarks by looking at the
self-consistent potentials and the ``force fields''
$F_{e}=-dV_{e}(z)/dz$ (short dashed curve) and $F_{g}=-dV_{g}(z)/dz$
(long dashed curve) -- note that $\Gamma_{vv'}^i \sim
\bra{v}F_{i}\ket{v}$, $i\in\{e,g,w\}$, -- in Fig.~\ref{fig:GaInAsPot3b}.
This figure was obtained for $V_b=1.2$ eV, but it does display the
general behavior for all quantities shown. The force field $F_{g}$
is essentially constant, except within the donor regions where the
wave functions are vanishingly small. Hence, the matrix element
$\bra{v}F_{g}\ket{v}$ [see Eqs.~\eqref{eq:etatot}--\eqref{eq:EtaB}]
is approximately linear in the external gate $V_{b}$. This explains
why the Rashba couplings $\alpha_v$ are strongly modulated by
external gates. This is even more so for $\alpha_1$, Fig.
\ref{fig:GaInAsA1a}, for which the structural and electronic
contributions cancel out. Looking at the wave functions $\psi_0$ and
$\psi_1$ and the force field $F_e=-dV_{e}/dz$ in
Fig.~\ref{fig:GaInAsPot3b}, we can see that the electronic Hartree
contribution ($\sim -\bra{v}F_{e}\ket{v}$) is almost zero (though
slightly negative) for the lowest subband and positive for the first
subband. The structural well contributions $\Gamma_{vv}^{w}$ [see
Eq.~\eqref{eq:EtaW}] to $\alpha_v$ are similar for both subbands,
though $|\Gamma_{00}^{w}|\ge |\Gamma_{11}^{w}|$, because the
nonzero biases ($V_{b}\neq 0$) cause the wave functions to shift
toward one side of the well [\textit{e.g.}, $V_b=1.2$ eV in
Fig.~\ref{fig:GaInAsPot3a}].

On the other hand, the contribution $\Gamma_{01}^g \sim
-\bra{0}F_{g}\ket{1}$ to the intersubband coupling $\eta$ is
essentially zero since the wave functions [$\psi_0$ and $\psi_1$ in
Fig.~\ref{fig:GaInAsPot3a}] are orthogonal and, again, $F_{g}$ is
constant. Hence $\eta$ is not as sensitive to the external gates as
the Rashba couplings. Most of the modulation of $\eta$ arises from
the electronic Hartree and structural contributions, which both have
the same sign and magnitude as shown in Fig.~\ref{fig:GaInAsE1a}.

\subsection{Double well}
\label{sec:InSb}

\subsubsection{Double-well parameters}
\label{sec:dwparameters}

Table~\ref{tab:bpInSb} shows the band
parameters\cite{jap89iv2001,prb72jmj2005} for the double quantum well
Al$_{0.4}$In$_{0.6}$Sb/InSb with one central barrier
InSb/Al$_{0.12}$In$_{0.88}$Sb. Hereafter we refer to this
heterostructure as InSb double well. The meaning of some of these
parameters (\textit{e.g.}, band offsets) can be seen in
Figs.~\ref{fig:bands4} and \ref{fig:qwell}.

\begin{table}[h]
  \caption{Relevant parameters\cite{jap89iv2001,prb72jmj2005} (at 1~K)
    for the InSb double well see Figs.~\ref{fig:bands4} and
    \ref{fig:qwell}). The width of the
    doping regions is $w=4$~nm and their densities are
    $\rho_{a}=\rho_{b}=3\times 10^{18}$~cm$^{-3}$. All energies are in
    eV and lengths in nm. The coefficient $\protect\eta_{H}$ is
    measured in nm$^{2}$ while $\eta_{w}$ and $\eta_{b}$ are measured
    in meV\,nm$^{2}$. The Dresselhaus constant $\beta{D}$ is measured
    in meV\,nm$^{3}$. The parameters in the last column are to
    the InSb binary compound.}
  \label{tab:bpInSb}
  \begin{ruledtabular}
    \begin{tabular}{rcl|rcl|rcl|rcl}
      $E_{w}$&=&$0.9922$ & $\Delta_{w}$&=&$0.6964$ & $w$&=&$4$ &
      $E_{P}$&=&$23.3$ \\ \hline
      $E_{g}$&=&$0.2350$ & $\Delta_{g}$&=&$0.8100$ & $\delta_{6}$&=&$0.6133$ &
      $\eem/m_{0}$&=&$0.0135$ \\ \hline
      $E_{b}$&=&$0.4477$ & $\Delta_{b}$&=&$0.7675$ & $\delta_{b6}$&=&$0.1723$ &
      $\epsilon_{r}$&=&$16.8$ \\ \hline
      $L\phantom{_{g}}$&=&$100$ & $L_{d}$&=&$65$ & $L_{b}$&=&$20$ &
      $L_{w}$&=&$50$ \\ \hline
      $\eta_{H}$&=&$0.2171$ & $\eta_{w}$&=&$0.7627$ & $\eta_{b}$&=&$5.0873$ &
      $\beta{D}$&=&$0.326$\\
    \end{tabular}
  \end{ruledtabular}
\end{table}

\subsubsection{SO couplings: double-well case}

\begin{figure}[h]
  \centering
  \subfigure[\label{fig:InSbCC1} $n_{T}$ constant.]
    {\resizebox{\imsizeC}{!}{\includegraphics{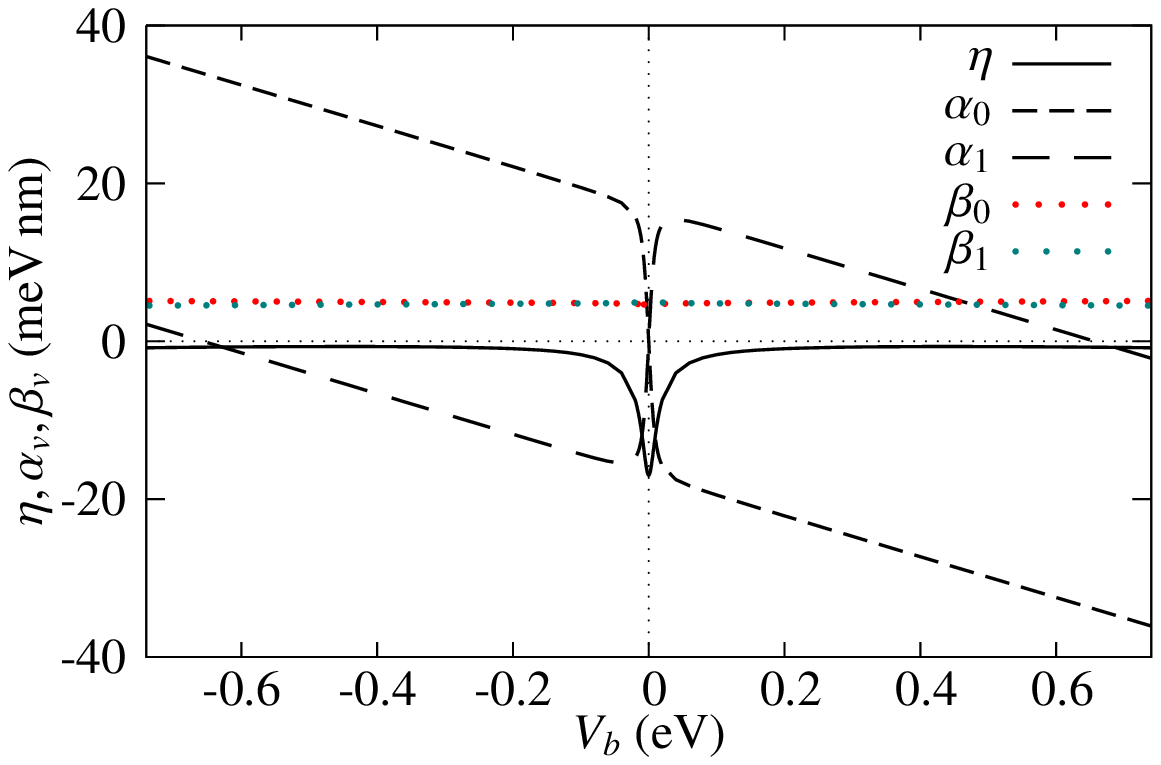}}}\\
  \subfigure[\label{fig:InSbCC2} $\mu$ constant.]
    {\resizebox{\imsizeC}{!}{\includegraphics{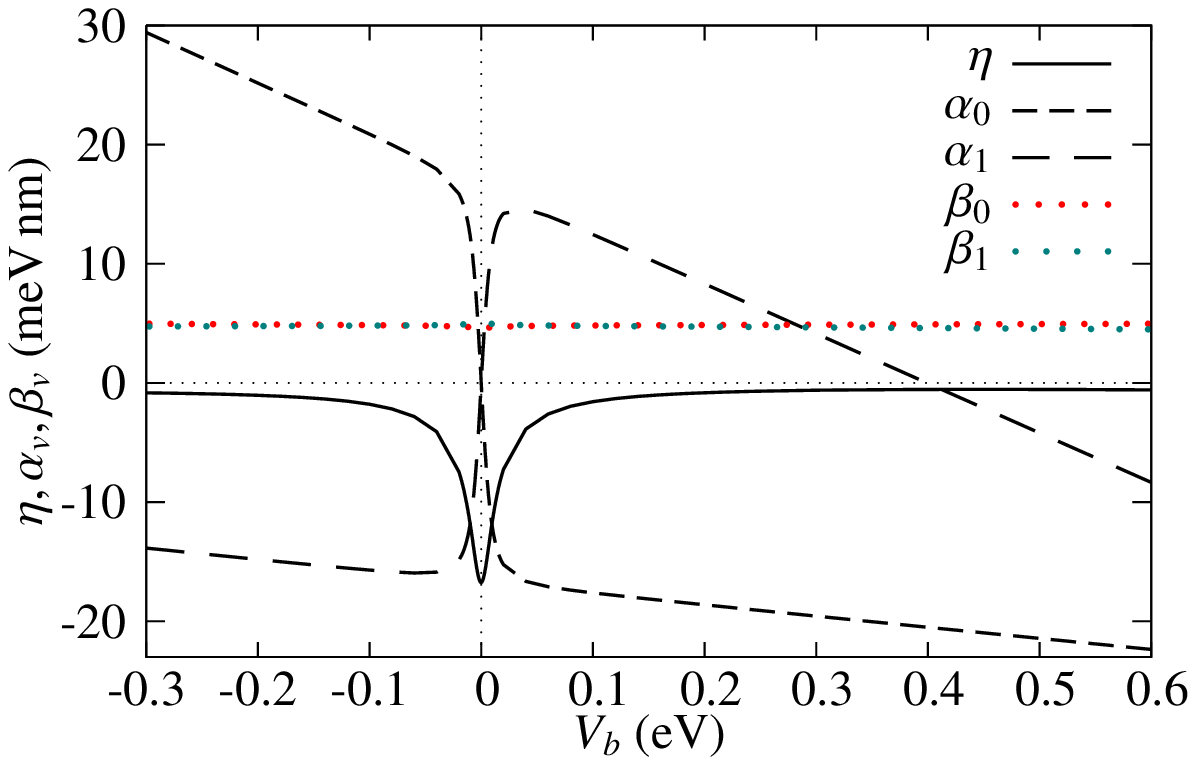}}}
    \par\vspace{-0.25cm}
    \caption{Rashba $\alpha$, Dresselhaus $\beta$ and the
      intersubband-induced $\eta$ SO couplings for a InSb double
      quantum well as functions of the right gate voltage $V_{b}$. In
      (a) the total electron density is kept constant at
      $n_{T}=10\times 10^{11}$~cm$^{-2}$ and in (b) the chemical
      potential is kept constant at $\mu=100$~meV (relative to to
      initial bottom well).}
    \label{fig:InSbCC}
\end{figure}

Figure~\ref{fig:InSbCC} shows the Rashba $\alpha$, Dresselhaus
$\beta$, and intersubband-induced $\eta$ SO couplings as functions
of the gate voltage $V_{b}$ (here again $V_{a}=0$) for both the
$n_T$-constant (a) and $\mu$-constant (b) models. We first discuss
the $n_T$-constant model [Fig.~\ref{fig:InSbCC1}]. Here the Rashba
couplings (dashed lines) are most sensitive to the external bias
$V_{b}$, being essentially the largest of all SO couplings for very
asymmetric structures (\textit{i.e.}, high biases). The Dresselhaus
couplings (dotted lines) are almost identical
($\beta_0\approx\beta_1$) and mostly independent of the external
gates. The SO coupling $\eta$ (solid line) is an even function of the
external gate $V_{b}$ and presents a ``resonant behavior'' around
the $V_{b}=V_{a}=0$~eV configuration, at which our sample is
symmetric. While the Rashba couplings are both zero at this
symmetric configuration, we note that they are odd functions of the
external gate (with $|\alpha_0| > |\alpha_1|$), have opposite signs
and abruptly change magnitudes around $V_b=0$ (over a 40~meV wide
region). For the $\mu$-constant model [Fig.~\ref{fig:InSbCC2}], a
similar picture as above also holds; note, however, that in contrast
to the $n_T$-constant model, in the $\mu$-constant case the positive
and negative bias configurations are not equivalent as they
correspond to the well having different numbers of electrons.

\begin{figure}[h]
  \centering
  \subfigure[\label{fig:InSbCC1d1} $\alpha_{v}$ and $\eta$ near $V_{b}=0$.]
    {\resizebox{\imsizeA}{!}{\includegraphics{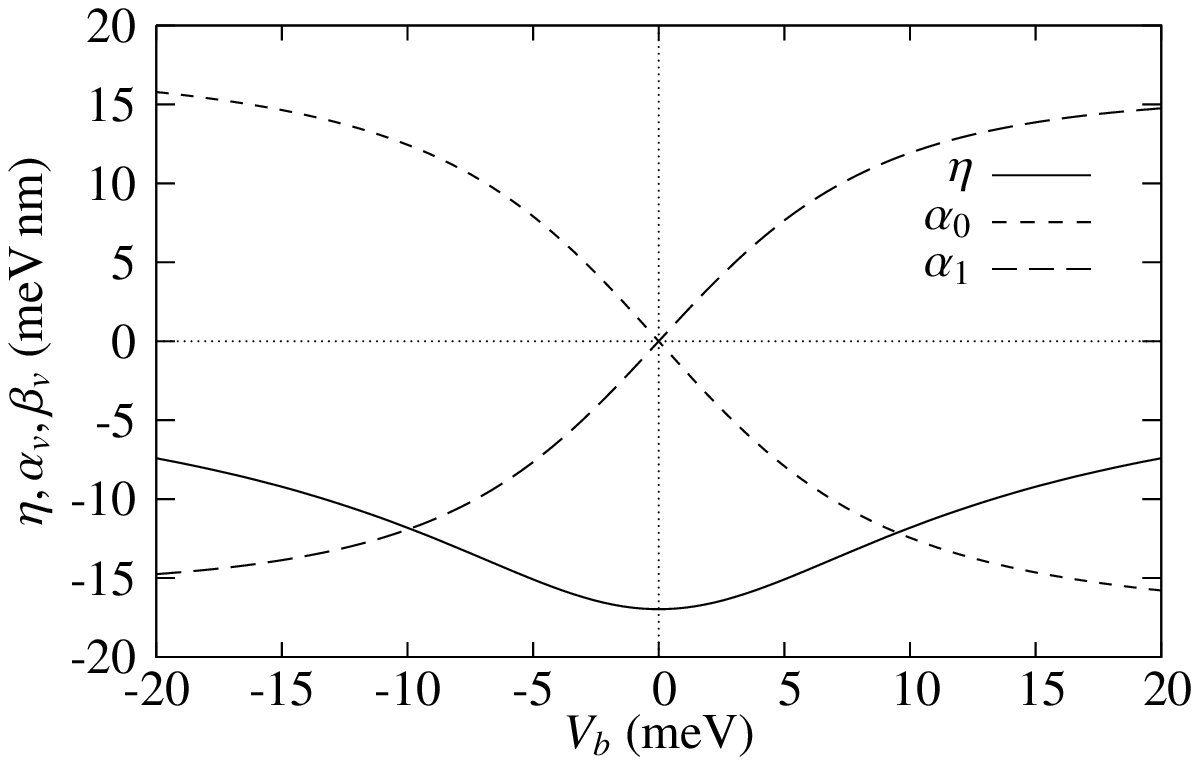}}}
  \subfigure[\label{fig:InSbCC1d2} $\beta_{v}$ near $V_{b}=0$.]
    {\resizebox{\imsizeA}{!}{\includegraphics{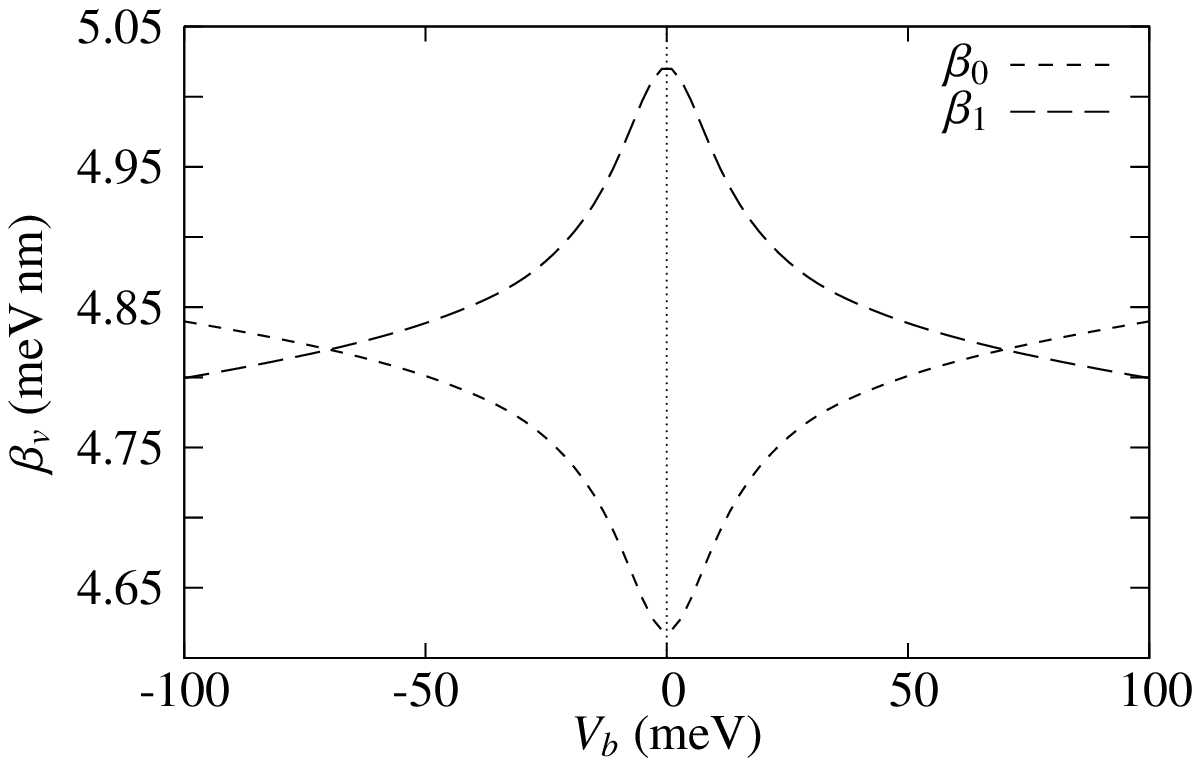}}}
  \par\vspace{-0.25cm}
  \caption{Rashba $\alpha$, intersubband-induced $\eta$ and
    Dresselhaus $\beta$ couplings vs $V_b$ about the symmetric
    configuration $V_{b}=V_{a}=0$~eV for the double InSb well in
    Fig.~\ref{fig:InSbCC1}.}
  \label{fig:InSbCC1d}
\end{figure}

For completeness we show in Fig.~\ref{fig:InSbCC1d} the behavior of
all coupling constants near the symmetric point $V_{b}=V_{a}=0$~eV for
the double well in Fig.~\ref{fig:InSbCC1}. Note that the Dresselhaus
couplings $\beta_0$ and $\beta_1$ present a (double) crossing over a
160~meV wide region [see Fig.~\ref{fig:InSbCC1d2}].  However, this is
a minor effect: note the change in the scale of the vertical axis.
While the resonant behavior of $\eta$ is accompanied by an enhancement
of about 10 in its magnitude [see Fig.~\ref{fig:InSbCC1}], we see no
substantial change in the magnitudes of the $\beta$'s near the
zero-bias case [\textit{cf.}  Figs.~\ref{fig:InSbCC1d1} and
\ref{fig:InSbCC1d2}].

\begin{figure}[h]
  \centering
  \subfigure[\label{fig:InSbCCeta1} $n_{T}$ constant.]
    {\resizebox{\imsizeA}{!}{\includegraphics{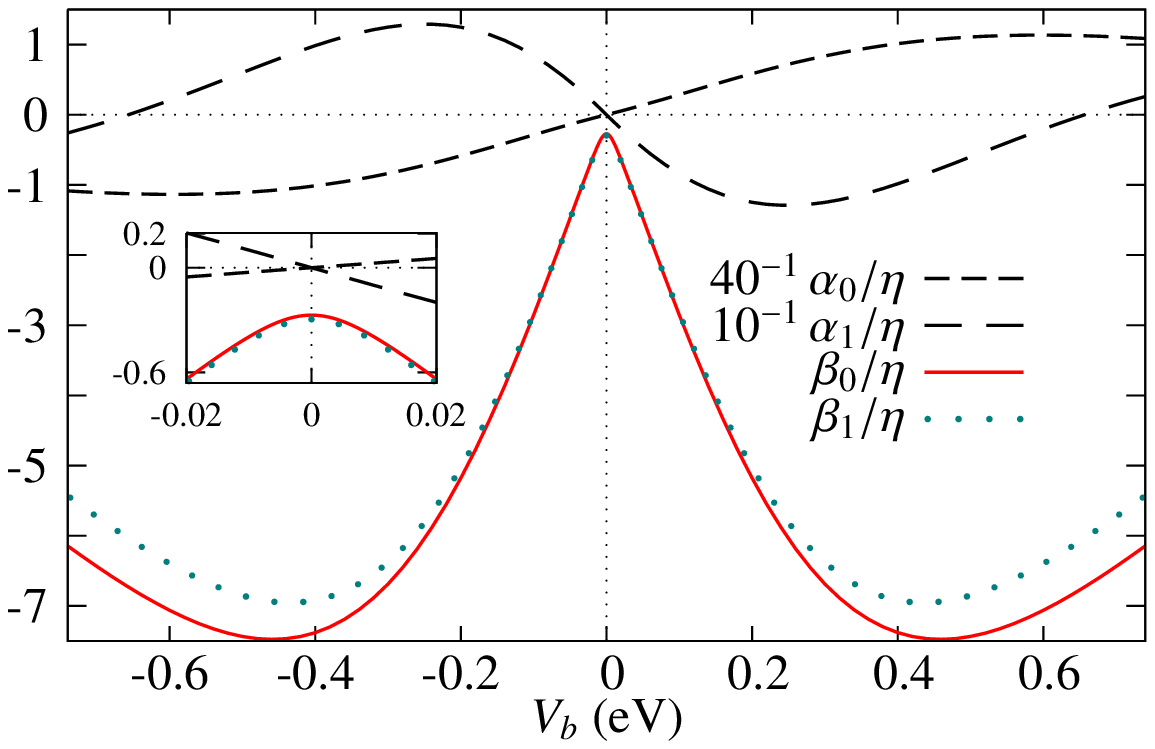}}}
  \subfigure[\label{fig:InSbCCeta2} $\mu$ constant.]
    {\resizebox{\imsizeA}{!}{\includegraphics{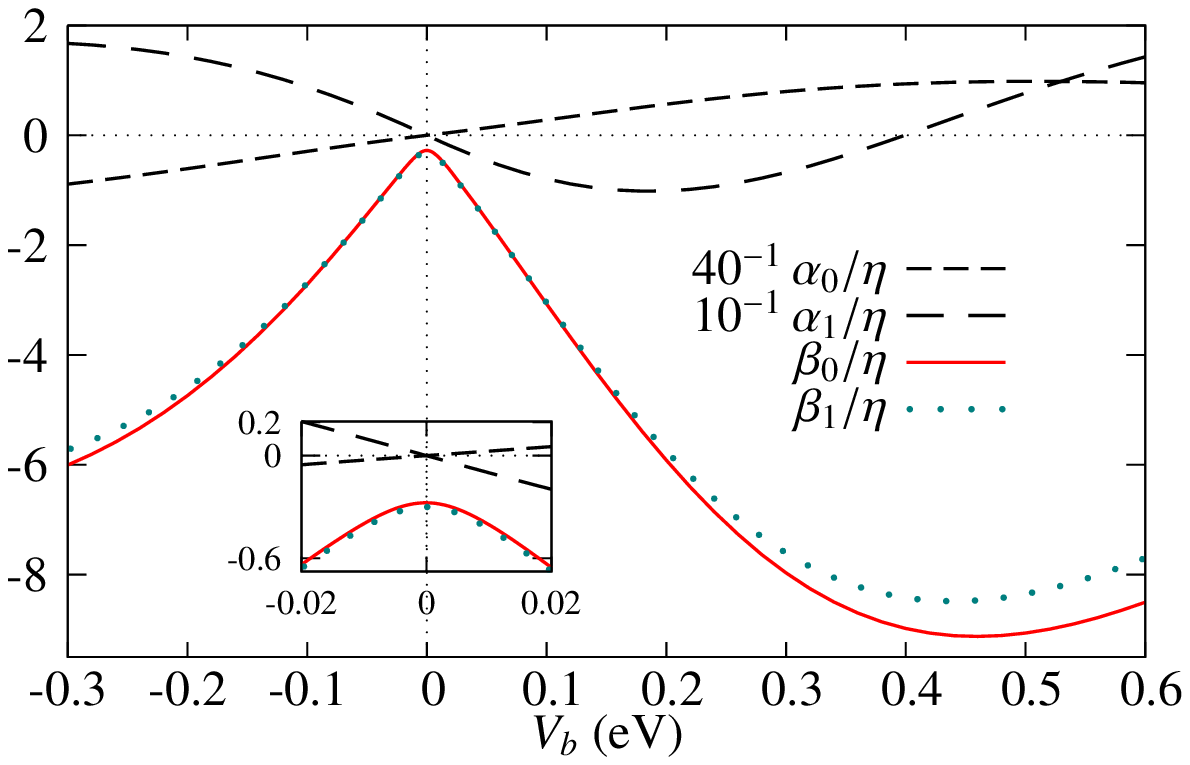}}}
  \par\vspace{-0.25cm}
  \caption{Ratios $\alpha_v/\eta$ and
    $\beta_v/\eta$ for the InSb double well in Fig.~\ref{fig:InSbCC}.
    The insets are blowups around $V_{b}=0$, the symmetric configuration.}
    \label{fig:InSbCCeta}
\end{figure}

The relative strengths of the Rashba and Dresselhaus coupling
constants to the intersubband-induced SO coupling are shown in
Fig.~\ref{fig:InSbCCeta}. The Rashba couplings have the largest
strengths (note the pre-factors in front of $\alpha_{v}/\eta$ in the
legends). In contrast to $\beta_{v}/\eta$, the linear behavior of
the Rashba ratios $\alpha_{v}/\eta$ near $V_{b}=0$ (see insets)
shows that $\alpha_{v}$ and $\eta$ undergo similar variations near
the symmetric configuration. As observed before, the
intersubband-induced coupling $\eta$ becomes important near
$V_{b}=0$ (Fig.~\ref{fig:InSbCC}).

\begin{figure}[ht]
  \centering
  \subfigure[\label{fig:InSbE1a} Contributions to $\eta$.]
    {\resizebox{\imsizeC}{!}{\includegraphics{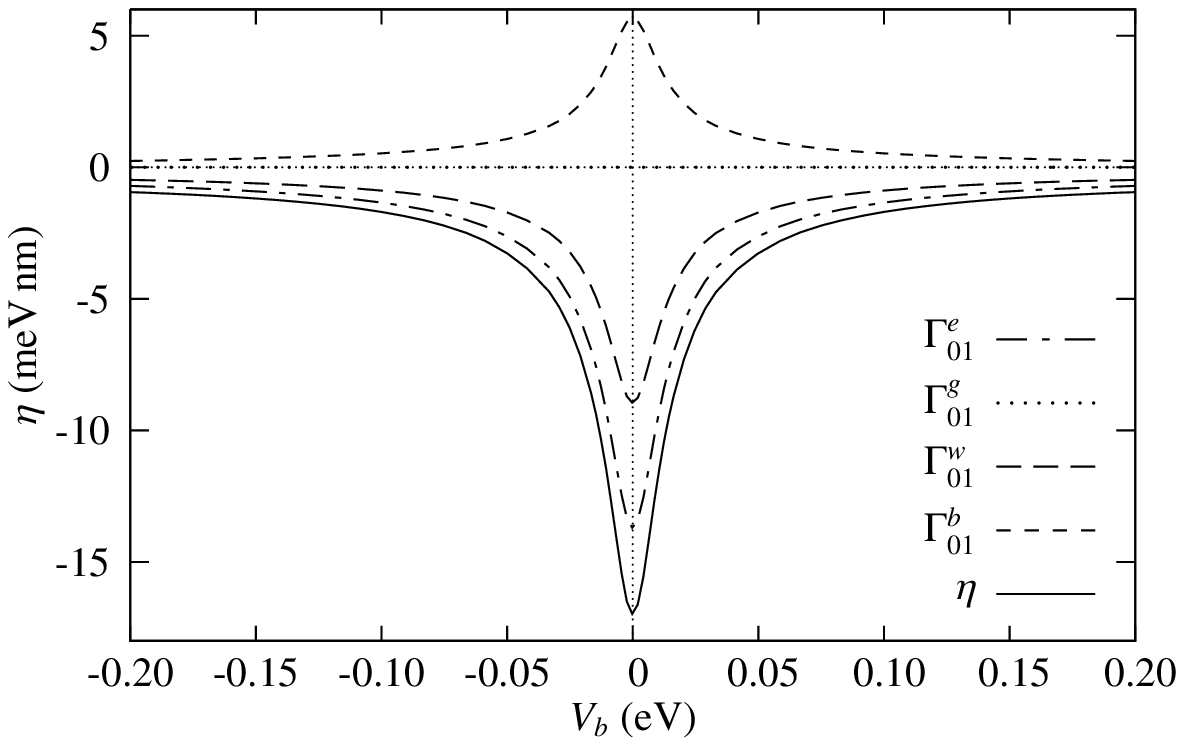}}}
  \subfigure[\label{fig:InSbA0a} Contributions to $\alpha_{0}$.]
    {\resizebox{\imsizeC}{!}{\includegraphics{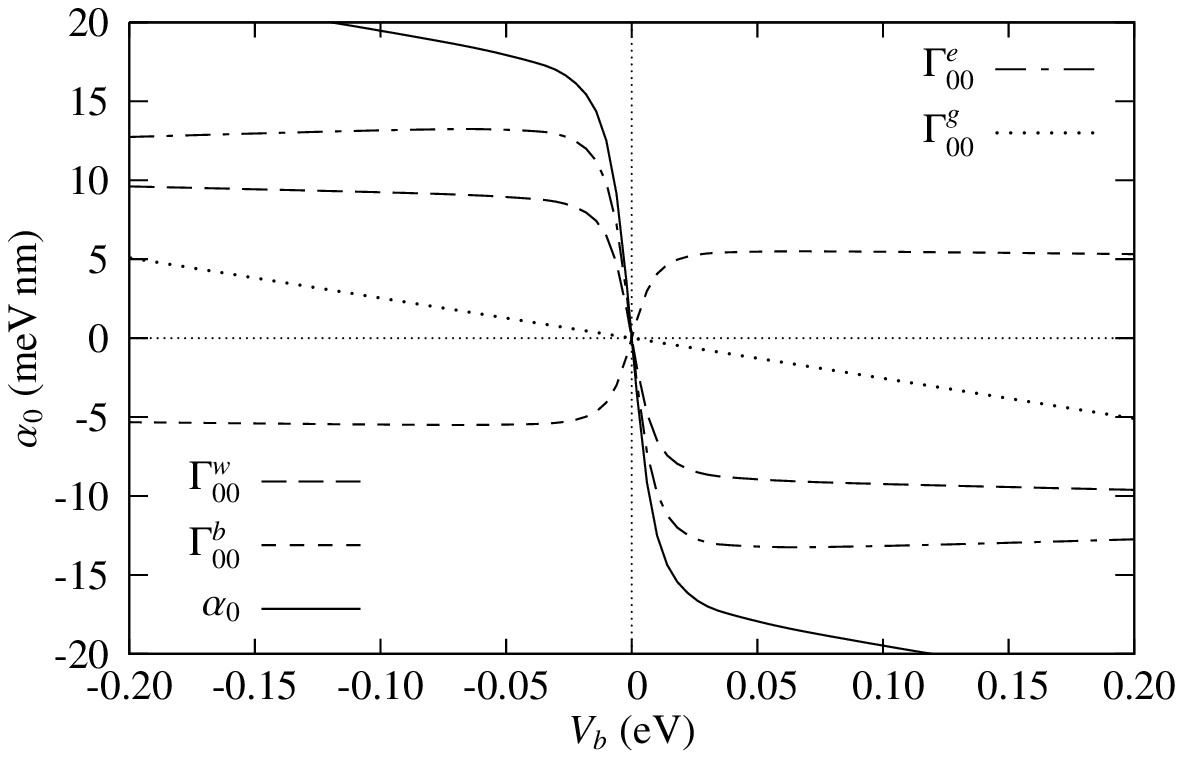}}}
  \subfigure[\label{fig:InSbA1a} Contributions to $\alpha_{1}$.]
    {\resizebox{\imsizeC}{!}{\includegraphics{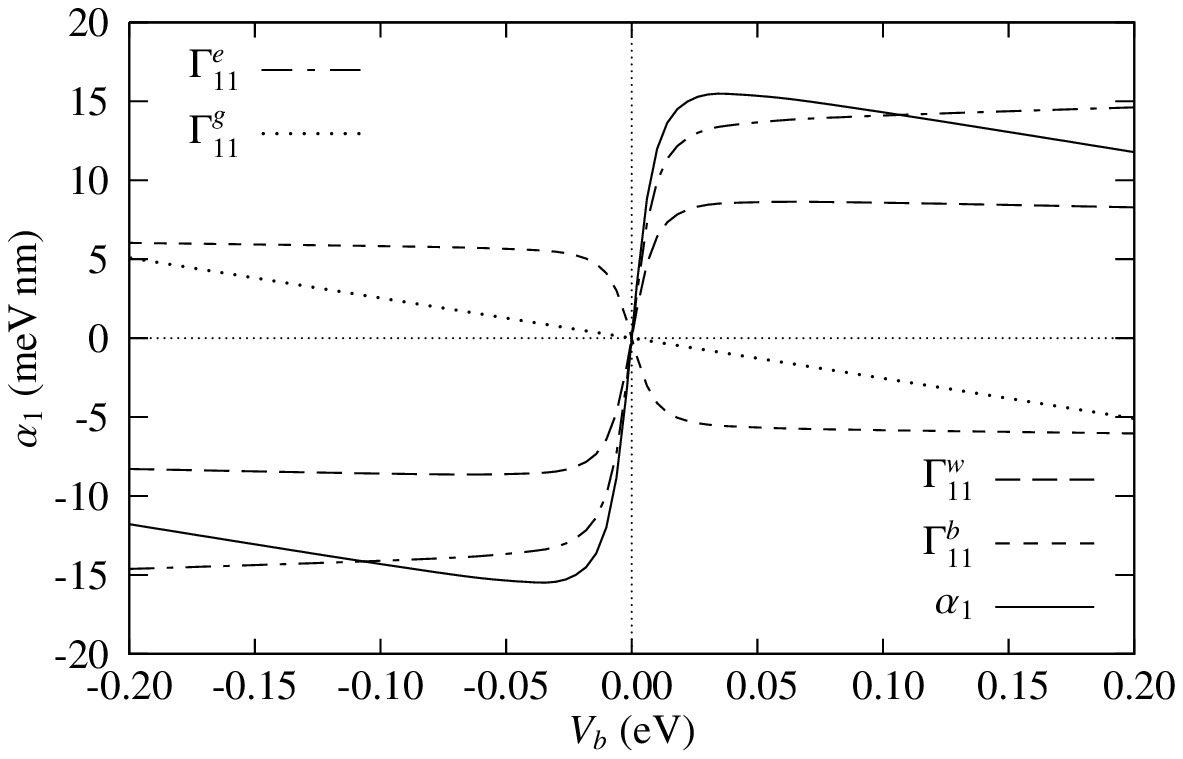}}}
  \par\vspace{-0.25cm}
  \caption{Different contributions to the SO
    couplings $\eta$, $\alpha_{0}$ and $\alpha_{1}$ for the InSb double
    well in the constant areal density model [see
    Fig.~\ref{fig:InSbCC1}] as functions of the external gate $V_{b}$
    ($V_{a}=0$). In the subfigures we show the contributions to the
    coupling constants coming from the areal electronic density,
    indicated by the superscript \textit{e}, and from the external gate +
    donor regions \textit{g}, and also from the structural potential, being
    \textit{w} for the well and \textit{b} for the central barrier.}
  \label{fig:InSbAEa}
\end{figure}

Figure~\ref{fig:InSbAEa} (similar to Fig.~\ref{fig:GaInAsAE} for the
single-well case) shows the several contributions to each of the SO
couplings $\eta$, $\alpha_0$ and $\alpha_1$ for the double-well
case. Here, in addition to the electronic Hartree, the gate (+
doping regions), and the well contributions, there is an additional
structural term arising from the central barrier (superscript
\textit{b}). A general feature in Figs.
\ref{fig:InSbE1a}--\ref{fig:InSbA1a} is that the structural
contributions (well and central barrier) almost cancel out because
they have opposite signs (see the curves with superscripts
\textit{w} and \textit{b}). These terms have opposite signs because
the derivatives $dh_w(z)/dz$ (well) and $dh_b(z)/dz$ (barrier),
which enter the coupling constants [see Eqs.~\eqref{eq:etatot},
\eqref{eq:EtaW}, and \eqref{eq:EtaB}], have opposite slopes. Similarly
to the single-well case [Fig.~\ref{fig:GaInAsE1a}] the contribution
of the external gates (which includes the doping regions) to the
intersubband SO coupling $\eta$ is vanishingly small [see the
$\Gamma_{01}^{g}$ curve in Fig.~\ref{fig:InSbE1a}].  Hence, $\eta$
is mostly due to the electronic Hartree contribution [curve
$\Gamma_{01}^{e}$ in Fig.~\ref{fig:InSbE1a}].  In addition, the gate
contribution to $\alpha_0$ and $\alpha_1$ for the InSb double well
is linear in $V_b$ as for the single-well case.  Hence, the Rashba
couplings $\alpha_0$ and $\alpha_1$ for the double InSb well are
essentially determined by the electronic (Hartree) contribution and
are modulated by the gate contribution. Summarizing: looking at
Fig.~\ref{fig:InSbAEa}, we can see that (i) the structural
contributions (well and barrier; dashed curves) almost cancel out,
and (ii) the external gate (dotted curves) modulates the Rashba
couplings $\alpha_{v}$; therefore, for the double well investigated
here (iii) most of the strength of these three coupling constants
($\eta$, $\alpha_{0}$, and $\alpha_{1}$) comes from the electronic
contribution (dot-dashed curves).

\begin{figure}[h]
  \centering
  \resizebox{\imsizeB}{!}{\includegraphics{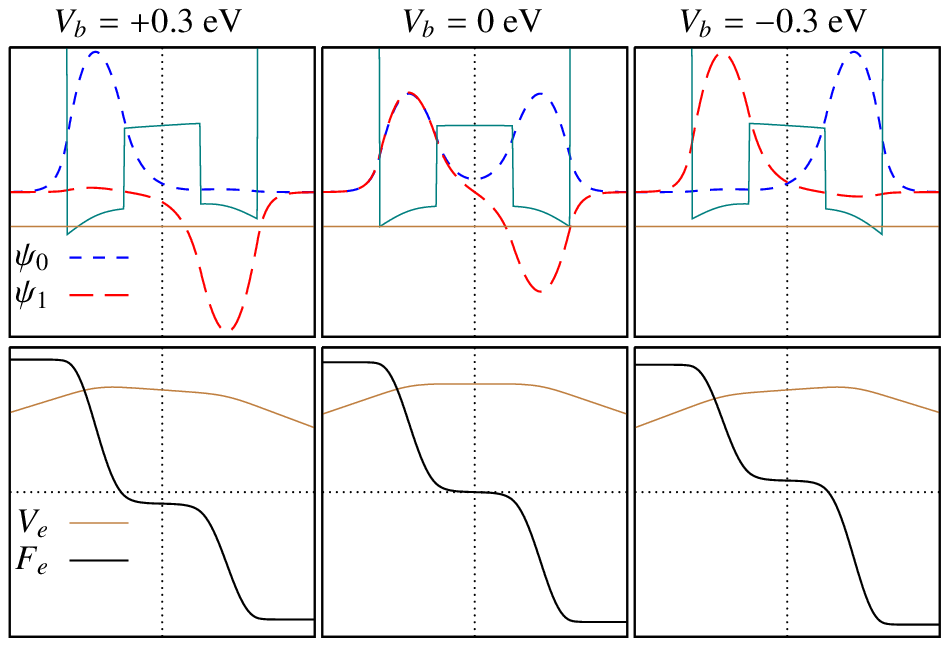}}\\
  \par\vspace{-0.15cm}
  \resizebox{\imsizeB}{!}{\includegraphics{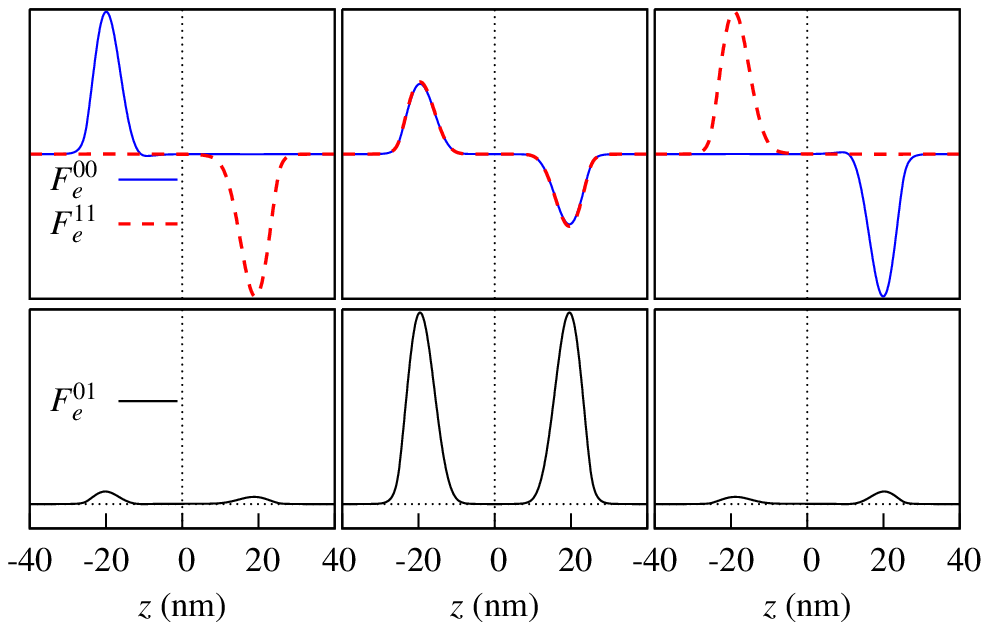}}
  \par\vspace{-0.25cm}
  \caption{Top row: Wave functions $\psi_{0}$ and $\psi_{1}$ for the
    InSb double well (Sec.~\ref{sec:dwparameters}) with external
    biases $V_{b}=+0.3$~eV, $V_{b}=0$~eV and $V_{b}=-0.3$~eV (1st, 2nd
    and 3rd columns, respectively).  The electronic Hartree
    contribution $V_e(z)$ to the potential energy and the
    corresponding force field $F_{e}=-dV_{e}/dz$ are shown on the
    second row. The third row shows the quantities
    $F_{e}^{00}(z)=\psi_{0}F_{e}\psi_{0}$ and
    $F_{e}^{11}(z)=\psi_{1}F_{e}\psi_{1}$ and the bottom row shows
    $F_{e}^{01}(z)=\psi_{0}F_{e}\psi_{1}$.  Here $n_{T}=10\times
    10^{11}$~cm$^{-2}$. The vertical dotted lines marks the center of
    the 20~nm wide AlInSb barrier within the 50~nm wide InSb well.}
  \label{fig:InSbPsis}
\end{figure}

It is instructive to investigate in more detail how the resonant
behavior in $\eta$ comes about, as well as the abrupt changes in the
Rashba couplings; see Fig.~\ref{fig:InSbCC}. This can be accomplished by
looking more closely at the self-consistent wave functions of the InSb
double well around the symmetric configuration ($V_{b}=0$). The top
row in Fig.~\ref{fig:InSbPsis} shows the self-consistent potential
profile of the double well and the normalized wave functions
$\psi_{0}$ (short dashed line) and $\psi_{1}$ (long dashed line) for
the lowest $v=0$ and for the first excited $v=1$ subbands at three
distinct gate voltages: $V_{b}=+0.3$~eV, $V_{b}=0$~eV and
$V_{b}=-0.3$~eV (left, center, and right columns, respectively). For
positive bias $\psi_{0}$ is mostly localized in the left well and
$\psi_{1}$ in the right well, while for negative biases this
configuration is reversed. The electronic Hartree contribution to the
potential energy $V_e$ and the corresponding force field
$F_{e}=-dV_{e}/dz$ are shown on the second row, thin and thick lines
respectively. Notice that $F_{e}$ is practically zero in the central
barrier region ($-10\leq z\leq 10$~nm) and has opposite signs within
the wells ($-25\leq z\leq -10$~nm and $10\leq z\leq 25$~nm). Hence the
quantities $F_{e}^{00}(z)=\psi_{0}(z) F_{e}\psi_{0}(z)$,
$F_{e}^{11}(z)=\psi_{1}(z) F_{e}\psi_{1}(z)$ and
$F_{e}^{01}(z)=\psi_{0}(z) F_{e}\psi_{1}(z)$ have the forms shown on
the third and fourth rows. The integral over $z$ of these quantities
defines the electronic Hartree contributions to the spin-orbit
couplings $\alpha_0$, $\alpha_1$, and $\eta$, \textit{i.e.},
$\Gamma_{00}^e \sim \bra{0}F_{e}\ket{0}$, $\Gamma_{11}^e \sim
\bra{1}F_{e}\ket{1}$, and $\Gamma_{01}^e \sim \bra{0}F_{e}\ket{1}$,
respectively. Since the electronic Hartree contributions dominate over
the others, see Figs. \ref{fig:InSbE1a}--\ref{fig:InSbA1a}, the
abrupt changes in the Rashba couplings and the resonant behavior of
$\eta$ around $V_b=0$ follow straightforwardly.

\subsubsection{Density anticrossings and effective masses}

\begin{figure}[h]
  \centering
  \includegraphics[scale=0.75]{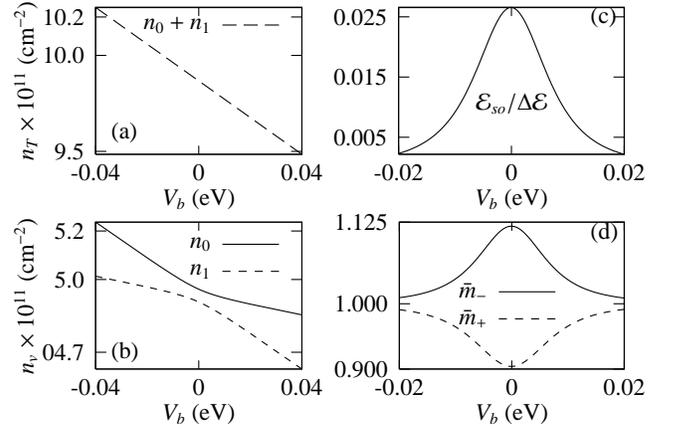}
  \par\vspace{-0.25cm}
  \caption{(a) Total electronic areal density $n_{T}=n_{0}+n_{1}$, (b) subband
areal densities $n_{0}$ and $n_{1}$, (c) ratio
$\mce_{so}/\Delta\mce$ and (d) effective mass ratios $\bar{m}_{\pm}
= \eem_{\pm}/\eem$ [Eq. \eqref{eq:SOEM}], as functions of the
external gate $V_{b}$ ($V_{a}=0$) in the $\mu$-constant model
($\mu=100$~meV).}
  \label{fig:InSbEso}
\end{figure}

Figures \ref{fig:InSbEso}(a) and \ref{fig:InSbEso}(b) show
anti-crossings of the areal densities $n_T$ for the InSb double well
near the symmetric configuration $V_{b}=0$ \cite{prb71rf2005}, where
the strength of the intersubband-induced SO coupling $\eta$ is the
strongest ($-16.7482$~meV\,nm) while the the energy difference between
the subband edges $\Delta\mce = \mce_{1}-\mce_{0}$ ($0.9353$~meV) is
the smallest. In accord with Eq.~\eqref{eq:RDk0}, we find an
appreciable change in the bulk effective mass $m^{\ast}$ near $\kpa=0$
\cite{saraga-2005}. The ratio $\mce_{so}/\Delta\mce$ [see
Eq.~\eqref{eq:SOEM}] is shown in Fig.~\ref{fig:InSbEso}(c) and the
ratio $\bar{m}_{\pm} = \eem_{\pm}/\eem$ in Fig.~\ref{fig:InSbEso}(d).
These intersubband-SO-induced changes in the effective masses
$\eem_{\pm}$ may have a sizable effect on the measured mobilities and
cyclotron frequencies in InSb wells.

\section{Summary}
\label{sec:fim}

Starting from the $8\times 8$ Kane model in heterostructures, we
have derived in some detail an effective electron Hamiltonian which
contains a new intersubband-induced SO interaction term which arises
in quantum wells with more than one quantized subband. Unlike the
usual Rashba SO term, the intersubband SO coupling here is non-zero
even for symmetric wells. For structurally asymmetric wells we have
also accounted for the Rashba-type SO interaction within each
subband.

We have also outlined the projection procedure (``folding down'') to
obtain quasi-2D Hamiltonians by integrating out the confined
variables. For two subbands in asymmetric wells we find a $4\times
4$ quasi-2D Hamiltonian resembling Rashba's, but containing three SO
couplings: the two Rashba couplings $\alpha_0$ and $\alpha_1$ and
the intersubband SO coupling $\eta$. For this two-subband case, we
have investigated thoroughly these three SO couplings for realistic
modulation-doped single and double wells. By performing a detailed
self-consistent calculation in which we solve both Poisson's and
Schr{\"{o}}dinger's equation iteratively, we have determined the
strengths of $\alpha_0$, $\alpha_1$, and $\eta$.

Each of these coupling strengths contains contributions arising from
(i) the potential-well (and barrier) offsets, (ii) the electronic
Hartree potential, and (iii) the external gate potential plus the
modulation doping potential. We have performed our simulations by
either keeping the areal electron density $n_T$ in the well fixed
($n_T$-constant model) or by keeping the chemical potential $\mu$
fixed ($\mu$-constant model). In the parameter range investigated,
both models give similar result for the calculated SO couplings.

For the single well investigated, $\alpha_0$ is mostly determined by
the external gate (+ modulation doping) contribution; with the
structural + electronic Hartree being about half of that of the
gate. On the other hand, $\alpha_1$ is essentially determined by the
external gate (+ modulation doping) contribution, since the
electronic Hartree and the structural contributions cancel out.
 The intersubband SO coupling $\eta$ is
essentially determined by the electronic Hartree potential and the
structural potential contributions (both of the same order); the
external gate (+ modulation doping) potential contribution to $\eta$
is nearly zero. Hence, while $\alpha_0$ and $\alpha_1$ can be
modulated by the external gate potential, $\eta$ is only slightly
influenced by it.

For double wells the SO couplings show more peculiar behaviors.  While
the Rashba couplings $\alpha_0$ and $\alpha_1$ abruptly change
magnitudes and signs around the symmetric configuration (zero external
bias), the intersubband-induced SO coupling presents a resonant
behavior being enhanced by a factor of 10 (with no sign change) around
this point. For the double well investigated the structural
contributions to $\alpha_0$, $\alpha_1$ and $\eta$, due to the
potential offsets of the edges of the well and the central barrier,
cancel out. In addition, the contribution of the external gate (+
doping region) to $\eta$ is vanishingly small (as for the single-well
case). Interestingly, the dominant contribution to all three SO
couplings $\alpha_0$, $\alpha_1$, and $\eta$ comes from the electronic
Hartree potential. However, this contribution is highly influenced by
the external gate, particularly around the symmetric configuration as
the electrons can easily localize in either well for slight (positive
or negative) changes in the gate.

Finally, we have also calculated the effective mass renormalization
due to the intersubband SO interaction (the Rashba-type interaction
does not produce a mass change). For the double well investigated,
we find that this mass renormalization is the largest ($\sim 10\%$)
around the symmetric potential configuration (zero external bias),
for which the splitting of the two subbands is the smallest. This
mass change can possibly have an effect on mobility and
cyclotron-resonance measurements.

\acknowledgments

We thank  G. J. Ferreira, H. J. P. Freire, and L. Viveiros for
useful discussions. This work was supported by the Swiss NSF, the
NCCR Nanoscience, JST ICORP, CNPq and FAPESP.

\appendix

\section{Self-consistent procedure}
\label{ap:sc-ap}

\subsection{Effective Schr{\"{o}}dinger equation}

The single-particle electron Hamiltonian $H_{QW}$ of our quantum wells
[Eq.~\eqref{eq:QWH})] is clearly separable. The transverse motion
(\textit{x,y}) is free while that along the \textit{z} direction is
confined by the quantum well. To solve the corresponding
Schr{\"{o}}dinger equation $H_{QW}\Psi_{\kpab v} (\bfl{r})=E_{\kpab
  v}\Psi_{\kpab v}(\bfl{r})$ we assume a wave function of the form
\begin{equation}
  \label{wf}
  \Psi_{\kpab v}(\bfl{r})=\langle \bfl{r}|\kpab v\rangle = \frac{1}{\sqrt{A}}
  \exp(i\kpab\cdot\rpab) \psi_{v}(z)
\end{equation}
($A$ is a normalizing area) which leads to the 1D Schr{\"{o}}dinger
equation
\begin{equation}
  \label{eq:schrod}
  \left( -\frac{\hbar^{2}}{2m}\frac{d^{2}}{dz^{2}} + V_{sc}(z)\right)
  \psi_{v}(z)= \left(E_{\kpab v} - \frac{\hbar^2\kpa^{2}}{2\eem}
  \right)= \mce_{v}\psi_{v}(z),
\end{equation}
from which we obtain the quantized energy levels $\mce_{v}$ and wave
functions $\psi_{v}(z)$. As we shall see, the subband structure of the
well $E_{\kpab v} = \mce_{v} + \hbar^2\kpa^{2}/2\eem$ and the
corresponding total wave function $\Psi_{\kpab v}(\bfl{r})$ will be
used (within a self-consistent procedure) to construct the electron
charge density, from which the corresponding Hartree potential can be
obtained via the Poisson equation.

As mentioned in Sec. \ref{subsec:expans} $V_{sc}$ in
Eq.~\eqref{eq:schrod} contains not only the structural confining
potential but also the ``Hartree contributions'' (i) the purely
electronic mean-field potential (electronic Hartree potential) and
(ii) the external gate potential plus the modulation doping potential.
Further down we discuss these contributions in detail.  Each of these
contributions is determined from a Poisson equation with an
appropriate charge distribution and boundary condition.

\subsubsection{Self-consistency}

Since the electronic charge distribution $\rho_e(z)$ ($\propto\!\!\!
\sum_v |\langle \bfl{r}|\kpab v\rangle|^2$) depends on the detailed
form of the several potentials (modulation doping, gates, and
electronic Hartree), and these, in turn, depend on $\rho_e(z)$, we
have to solve the problem self-consistently. The standard procedure
is as follows: (i) to solve Eq. \eqref{eq:schrod}) with an initial guess for
$V_{sc}$ which we take to be just the structural potential plus the
external gates and modulation doping potential [\textit{i.e.}, in the
first run we do not include the electronic Hartree potential
$V_e(z)$]; (ii) to construct the electronic charge density $\rho_e(z)$
[from the eigenfunctions obtained in step (i)] and the corresponding
$V_e(z)$ via Poisson equation; and (iii) to solve again the
Schr{\"{o}}dinger equation with the new $V_{sc}$, which in this new
iteration includes $V_e(z)$ (as well as the other potentials: gates,
modulation doping, and structural confinement). We repeat this process
until convergence is attained.

\subsubsection{Numerics}

We use the sixth-order Numerov method to solve the Schr{\"{o}}dinger
equation.\cite{rasmn84bvn1924,jcp1jmb1967,ajp40pcs1972,cma0rpa2001}
Poisson equation (see Sec.~\ref{apx:poisson}) is solved via a
semi-analytical Numerov method.\cite{sub2007eb1} All numerical
integrations are performed using a Gaussian integration
method.\cite{abramowitz1964} In our numerical implementation we use
the dimensionless form of Eq.~\eqref{eq:schrod},
\begin{equation}
  \label{eq:aschrod}
  \frac{d^{2}\tilde{\psi}_{v}}{d\tilde{z}^{2}} = \tilde{V}_{v}\tilde{\psi}_{v},
  \quad \tilde{\psi}_{v}=\psi_{v}(\tilde{z}),\quad \tilde{z}=\frac{z}{l},
\end{equation}
where
\begin{equation}
  \label{eq:eps1}
  \tilde{V}_{v}=\frac{2\pi}{\varepsilon_{1}}
  \bigl[V_{sc}(\tilde{z})-\mce_{v}\bigr], \quad
  \varepsilon_{1}=\frac{\pi\hbar^{2}}{\eem l^{2}}.
\end{equation}
We choose $l=1$~nm as our length unit and $\varepsilon_{1}$ as the
relevant energy scale.

\subsection{Poisson equations for the electronic and gate plus
  modulation doping potentials}
\label{apx:poisson}

\begin{figure}[h]
  \centering
  \resizebox{\imsizeB}{!}{\includegraphics{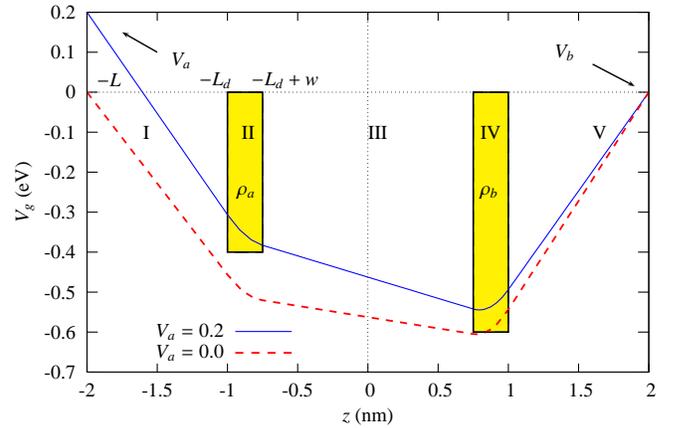}}
  \par\vspace{-0.25cm}
  \caption{Schematic representation of doping layers
    of width $w$ and densities $\rho_{a}$ and $\rho_{b}$ plus the
    external gates $V_{a}$ and $V_{b}$. By varying the external gates
    (we usually keep $V_a=0$ and vary $V_b$) we can alter the spatial
    symmetry of our quantum wells, Fig.~\ref{fig:qwell}.  The curves
    illustrate the calculated gate + modulation doping potential
    $V_g(z)$ for $V_a=V_b$ (dashed line) and $V_a>V_b$ (solid line),
    both with $\rho_a>\rho_b$.}
  \label{fig:layers}
\end{figure}

The self-consistent electronic potential energy
$V_{sc}(z)=-e\phi_{sc}(z)$ can be split in two parts,
$V_{sc}(z)=V_{wb}(z) + V_{H}(z)$. $V_{wb}(z)=V_{w}(z)+V_{b}(z)$
described the structural quantum-well potential. The ``Hartree''
contribution $V_{H}(z)=V_{e}(z)+V_{g}(z)$ arises from the electronic
charge density and from the external gates plus the modulation
doping regions (symmetrically located around the well; see
Fig.~\ref{fig:layers}). Figure \ref{fig:layers} also shows the
Dirichlet boundary conditions $V_{a}$ and $V_{b}$, which are in fact
the external gates at the end points $\pm L$ of our system.

\subsubsection{Gate+modulation doping potential}

We can write separate Poisson equations for $V_{g}(z)$ and $V_e(z)$
as these arise from distinct charge densities. For $V_g(z)$ we have
(see Fig.~\ref{fig:layers})
\begin{equation}
  \label{eq:PoissonD}
  \frac{d^{2}}{dz^{2}} V_{g} = \frac{e^{2}}{\epsilon_{r}\epsilon_{0}}
  \begin{cases}
    0, & \text{(I)}: -L \leq z \leq -L_{d}, \\
    \rho_{a}, & \text{(II)}: -L_{d}\leq z\leq -L_{d}+w, \\
    0, & \text{(III)}: -L_{d}+w \leq z \leq L_{d}-w, \\
    \rho_{b}, & \text{(IV)}: L_{d}-w\leq z\leq L_{d}, \\
    0, & \text{(V)}: L_{d} \leq z \leq L,
  \end{cases}
\end{equation}
where $\epsilon_{0}$ is the permittivity, $\epsilon_{r}$ is the
dielectric constant\cite{prb27gas1983,lide2004} and $\rho_{a,b}$ are
the doping densities.  From the continuity of $V_{g}(z)$ and its first
derivative and assuming the Dirichlet boundary conditions
$V_{g}(-L)=V_{a}$ and $V_{g}(+L)=V_{b}$, we find
\begin{equation}
  \label{eq:Vd}
  V_{g} =
  \begin{cases}
    c_{1}z+c_{2},\; \text{(I)}: -L \leq z \leq -L_{d}, \\
    \frac{1}{2}Az^{2}+c_{3}z+c_{4},\; \text{(II)}: -L_{d}\leq z\leq
    -L_{d}+w, \\
    c_{5}z+c_{6},\; \text{(III)}: -L_{d}+w \leq z \leq L_{d}-w, \\
    \frac{1}{2}Bz^{2}+c_{7}z+c_{8},\; \text{(IV)}: L_{d}-w\leq z\leq L_{d}, \\
    c_{9}z+c_{10},\; \text{(V)}: L_{d} \leq z \leq L,
  \end{cases}
\end{equation}
with
\begin{equation}
  \label{eq:ab}
  A = \frac{e^{2}\rho_{a}}{\epsilon_{r}\epsilon_{0}},\quad
  B = \frac{e^{2}\rho_{b}}{\epsilon_{r}\epsilon_{0}},
\end{equation}
where the constants $c_{i}$ are given in Appendix~\ref{apx:cis}.
Figure \ref{fig:layers} shows two solutions of Eq.~\eqref{eq:PoissonD},
both having $\rho_{a}>\rho_{b}$ and $V_a=V_b$ (dashed line) and
$V_a>V_b$.

\subsubsection{Electronic Hartree potential}

The electronic Hartree contribution $V_{e}(z)$ is determined from
\begin{equation}
  \label{eq:PoissonE}
  \frac{d^{2}}{dz^{2}} V_{e}(z) =
  -\frac{e}{\epsilon_{r}\epsilon_{0}} \rho_{e}(z),
\end{equation}
with (including spin)
\begin{equation}
  \label{eq:rhoe}
  \rho_{e}(z) = \frac{2e}{A}\sum_{v,\kpa}|\psi_{v}(z)|^{2}f(E_{\kpa v}) =
  \frac{e\eem}{\pi\hbar^{2}}\,k_{B}T \lambda_{e}(z),
\end{equation}
where
\begin{equation}
  \label{eq:le}
  \lambda_{e}(z) = \sum_{v}|\psi_{v}(z)|^{2}
  \ln\bigl[1+\e^{\beta(\mu-\mce_{v})/k_{B}T}\bigr],
\end{equation}
and
\begin{equation}
  \label{eq:ffermi}
  f(E_{\kpa v}) = \frac{1}{1+\e^{(E_{\kpa v}-\mu)/k_{B}T}},\quad
  E_{\kpa v}=\frac{\hbar^{2}\kpa^{2}}{2\eem} + \mce_{v}.
\end{equation}
We solve Eq.~\eqref{eq:PoissonE} for  $V_{e}(z)$ using an accurate Numerov
scheme~\cite{sub2007eb1} with the Dirichlet boundary conditions
$V_{e}(\pm L)=0$. Similarly to the Schr{\"{o}}dinger equation in
Eq.~\eqref{eq:aschrod}), we find it convenient here to write the Poisson
equation \eqref{eq:PoissonE} in a dimensionless form
\begin{equation}
  \label{eq:PoissonE2}
  \frac{d^{2}}{d\ti{z}^{2}} \ti{V}_{e} = -\ti{\lambda_{e}},\quad
  \ti{\lambda_{e}} = \frac{k_{B}T}{\varepsilon_{1}}\,
  l\lambda_{e}(\ti{z}),\quad
  \varepsilon_{2}\ti{V}_{e}=V_{e}(\ti{z}),
\end{equation}
where $\varepsilon_{1}$ is the energy scale given in
Eq.~\eqref{eq:eps1} and
\begin{equation}
  \label{eq:eps2}
  \varepsilon_{2} = \frac{e^{2}}{\epsilon_{r}\epsilon_{0}l}.
\end{equation}

\subsubsection{Electron density and chemical potential}

From the total electronic charge
\begin{equation}
  \int\! dV\,\rho_{e}(z)=en_{T}A
\end{equation}
we can straightforwardly [using Eq.~\eqref{eq:rhoe})] obtain the total
areal concentration of electrons
\begin{equation}
  \label{eq:n0}
  n_{T} = \sum_{v}n_{v},
\end{equation}
with the $n_v$'s denoting the subband occupations
\begin{equation}
  \label{eq:nv}
 n_{v}=\frac{\eem}{\pi\hbar^{2}}\,
  k_{B}T\,\ln\left[1+\e^{(\mu-\mce_{v})/k_{B}T}\right].
\end{equation}

When $n_{T}$ is fixed (\textit{i.e.}, the $n_T$-constant model), we can
determine the chemical potential $\mu$ from Eq.~\eqref{eq:n0},
\begin{equation}
  \label{eq:n0b}
  \frac{\pi\hbar^{2}n_T}{\eem k_B T}=
  \sum_{v}\ln\left[1+\e^{(\mu-\mce_{v})/k_{B}T}\right].
\end{equation}

\section{Coefficients $c_i$'s}
\label{apx:cis}

Using the continuity of $V_g$ and its first derivative together with
the (Dirichlet) boundary conditions at the end points $V_g(-L)=V_a$
and $V_g(L)=V_b$, we can determine the coefficients $c_{i}$'s
appearing in Eq.~\eqref{eq:Vd}. In the regions \textit{I} and
\textit{V} we find
\begin{align}
  c_{1}  &= -\frac{2L_{d}-w}{2L}\,wC_{-} - wC_{+} -
    \frac{V_{-}}{L}, \\
  c_{2}  &= -\frac{1}{2}(2L_{d}-w)\,wC_{-} - L\,wC_{+} + V_{+}, \\
  c_{9}  &= -\frac{2L_{d}-w}{2L}\,wC_{-} + wC_{+} -
    \frac{V_{-}}{L}, \\
  c_{10} &= +\frac{1}{2}(2L_{d}-w)\,wC_{-} - L\,wC_{+} + V_{+},
\end{align}
with
\begin{equation}
  C_{\pm}=\frac{1}{2}(A\pm B), \quad
  V_{\pm}=\frac{1}{2}(V_{a}\pm V_{b}),
\end{equation}
and $A$ and $B$ defined in Eq.~\eqref{eq:ab}.  In the modulation doping
regions \textit{II} and \textit{IV}, we have
\begin{align}
  c_{3} &= \frac{w^{2}-2wL_{d}+2LL_{d}}{2L}\, C_{-} +
  (L_{d}-w)C_{+} - \frac{V_{-}}{L}, \\
  c_{4} &= +\frac{1}{2}(L_{d}-w)^{2}\, C_{-} +
  \frac{1}{2}(L_{d}^{2}-2wL)C_{+} + V_{+}, \\
  c_{7} &= \frac{w^{2}-2wL_{d}+2LL_{d}}{2L}\, C_{-} -
  (L_{d}-w)C_{+} - \frac{V_{-}}{L}, \\
  c_{8} &= -\frac{1}{2}(L_{d}-w)^{2}\, C_{-} +
  \frac{1}{2}(L_{d}^{2}-2wL)C_{+} + V_{+}.
\end{align}
In the central region \textit{III}, we have
\begin{align}
  c_{5} &= +\frac{2L-2L_{d}+w}{2L}\, wC_{-} - \frac{V_{-}}{L}, \\
  c_{6} &= -\frac{1}{2}(2L-2L_{d}+w)\, wC_{+} + V_{+}.
\end{align}

\bibliography{articles,books}

\begin{thebibliography}{44}
\bibitem[{neg()}]{neglect-terms}
\bibinfo{note}{Here we neglect the Darwin and all higher-order terms and focus
  on only the Rashba-like contributions.}
\bibitem[{kan()}]{kane}
\bibinfo{note}{The Kane effective mass in Eq.~\eqref{eq:Mtrans0} neglects
  corrections from higher bands; here we use $\eem$ as a parameter determined
  from experiment.}
\end{thebibliography}

\end{document}